\documentclass[prd,nofootinbib,superscriptaddress]{revtex4}
\usepackage{amsmath,amssymb,graphicx}

\renewcommand{\vec}[1]{\boldsymbol{#1}}
\newcommand{\dif}{\mathrm{d}}
\newcommand{\xB}{x_{\rm Bj}}
\newcommand{\chisq}{\chi^2/\mathrm{d.o.f.}}
\newcommand{\Pom}{{\mathbb{P}}}

\begin{document}

\title{Impact parameter dependent colour glass condensate dipole model}
\author{G.~Watt}
\affiliation{Department of Physics \& Astronomy, University College London, WC1E 6BT, UK}
\author{H.~Kowalski}
\affiliation{Deutsches Elektronen-Synchrotron DESY, 22607 Hamburg, Germany}

\begin{abstract}
  We show that the colour glass condensate dipole model of Iancu, Itakura and Munier, improved to include the impact parameter dependence, gives a good fit to the total $\gamma^*p$ cross section measured at HERA if the anomalous dimension at the saturation scale, $\gamma_s$, is treated as a free parameter.  We find that the optimum value of $\gamma_s=0.46$ is close to the value determined from numerical solution of the Balitsky--Kovchegov equation.  The impact parameter dependent saturation scale is generally less than 0.5 GeV$^2$ in the HERA kinematic regime for the most relevant impact parameters $b\sim 2$--$3$ GeV$^{-1}$.  We compare predictions of the model to data on the longitudinal and heavy flavour structure functions, exclusive diffractive vector meson production and deeply virtual Compton scattering at HERA.  The model is found to be deficient for observables sensitive to moderately small dipole sizes, where an alternative model with explicit DGLAP evolution performs better.  The energy dependence of exclusive diffractive processes is shown to provide an important discriminator between different dipole model cross sections.
\end{abstract}

\maketitle

\section{Introduction} \label{sec:introduction}

The colour dipole model has proven to be very successful in describing a wide variety of small-$x$ inclusive and diffractive processes at HERA.  In particular, it is commonly used in determinations of the saturation scale, that is, the $x$-dependent momentum scale at which nonlinear effects start to become important.  The total cross section for $\gamma^* p$ scattering,
\begin{equation}
  \sigma^{\gamma^*p}_{\rm tot} = \sigma_T^{\gamma^* p} + \sigma_L^{\gamma^* p} = \frac{4\pi^2\alpha_{\rm em}}{Q^2}\,F_2,
\end{equation}
is obtained by combining the light-cone wave functions for the virtual photon to fluctuate into a $q\bar{q}$ pair with the dipole cross section for the $q\bar{q}$ pair to scatter elastically off the proton:
\begin{equation}
  \sigma^{\gamma^* p}_{T,L} = \sum_f \int\!\dif^2\vec{r} \int_0^1\!\frac{\dif z}{4\pi}(\Psi^{*}\Psi)_{T,L}^f\,\int\!\dif^2\vec{b}\;\frac{\dif\sigma_{q\bar q}}{\dif^2\vec{b}}.
  \label{eq:siggp}
\end{equation}
Here, $f$ is the flavour of the $q\bar{q}$ pair, $z$ is the fraction of the photon's light-cone momentum carried by the quark, $r=|\vec{r}|$ is the transverse size of the $q\bar{q}$ dipole, while $\vec{b}$ is the impact parameter, that is, $b$ is the transverse distance from the centre of the proton to the centre of mass of the $q\bar{q}$ dipole.  The squared photon wave functions, $(\Psi^{*}\Psi)_{T,L}^f$, are given explicitly in Ref.~\cite{Kowalski:2006hc}.

A popular parameterisation for the $b$-integrated dipole cross section is the ``saturation model'' due to Golec-Biernat and W\"usthoff (GBW) \cite{Golec-Biernat:1998js,Golec-Biernat:1999qd}:
\begin{equation}
  \sigma_{q\bar q} \equiv \int\!\dif^2\vec{b}\;\frac{\dif\sigma_{q\bar q}}{\dif^2\vec{b}} = \sigma_0\left(1-\mathrm{e}^{-r^2Q_s^2(x)/4}\right),
  \label{eq:sigGBW}
\end{equation}
where $\sigma_0$ is a constant and $Q_s^2(x) = (x_0/x)^{\lambda}$ GeV$^2$.  The parameters $\sigma_0=29$ mb, $\lambda=0.28$ and $x_0=4\times 10^{-5}$ were obtained from a fit, including charm quarks, to inclusive deep-inelastic scattering (DIS) data \cite{Golec-Biernat:1998js}.  However, the GBW model does not give a good fit to recent DIS data \cite{Kowalski:2006hc}.

Many improvements to the seminal but simple GBW parameterisation \eqref{eq:sigGBW} of the dipole cross section have been proposed (see Ref.~\cite{Kowalski:2006hc} for a more comprehensive review and further references).  The GBW model was modified to include DGLAP evolution in the Bartels--Golec-Biernat--Kowalski (BGBK) model \cite{Bartels:2002cj} (extended to include heavy quarks in Ref.~\cite{Golec-Biernat:2006ba}).  An alternative colour glass condensate (CGC) model inspired by the Balitsky--Kovchegov (BK) equation\footnote{It has recently been shown \cite{Dumitru:2007ew} that running coupling effects strongly suppress the effect of ``Pomeron loops'' such that the BK equation should be sufficient for phenomenological studies.} \cite{Balitsky:1995ub,Kovchegov:1999yj,Kovchegov:1999ua} was proposed by Iancu, Itakura and Munier \cite{Iancu:2003ge} and extended to include charm quarks in Ref.~\cite{Kowalski:2006hc}.  The introduction of charm quarks in the CGC model led to a dramatic decrease in the saturation scale \cite{Kowalski:2006hc}.  However, it has recently been shown by Soyez \cite{Soyez:2007kg} that allowing the anomalous dimension at the saturation scale to increase from the fixed value of $\gamma_s = 0.63$ assumed in Refs.~\cite{Iancu:2003ge,Kowalski:2006hc} to a higher value of $\gamma_s = 0.74$ slightly improves the fit to the $F_2$ data and gives a larger saturation scale.  On the other hand, this value of $\gamma_s$ seems to be inconsistent with the value of $\gamma_s\simeq 0.44$ recently obtained from numerical solution of the BK equation by Boer, Utermann and Wessels \cite{Boer:2007wf}.

The BGBK \cite{Bartels:2002cj} and CGC \cite{Iancu:2003ge} models considered only the dipole cross section integrated over the impact parameter $b$.  However, the gluon density is larger in the centre of the proton ($b=0$) than at the typical impact parameters $b\sim 2$--$3$ GeV$^{-1}$ probed in the total $\gamma^*p$ cross section.  Therefore, any serious determination of the saturation scale at HERA should consider the impact parameter dependence of the dipole cross section.  The $b$ dependence is also necessary to describe the slope of $t$ distributions of diffractive processes at HERA, which in turn fix the normalisation of the $b$-integrated dipole cross section.  Therefore, in the context of dipole models, the analysis of inclusive HERA data should not be considered in isolation to diffractive HERA data.

The BGBK model was modified to include the impact parameter dependence in Refs.~\cite{Kowalski:2006hc,Kowalski:2003hm}, denoted by the ``b-Sat'' model, and was found to give a good description of both $F_2$ data and exclusive diffractive processes.  The CGC model was also modified to include the impact parameter dependence in Ref.~\cite{Kowalski:2006hc}, denoted by the ``b-CGC'' model, but it was not possible to get a good description of the $F_2$ data with a fixed value of $\gamma_s = 0.63$.  It is therefore interesting to see if a good fit with the b-CGC model can be obtained when $\gamma_s$ is allowed to vary and to compare the optimum value obtained with the expected value of 0.44 obtained from numerical solution of the BK equation \cite{Boer:2007wf}.

In Sec.~\ref{sec:cgc} we recall the original CGC model \cite{Iancu:2003ge} and investigate the dependence on $\gamma_s$ of the quality of the fit to HERA $F_2$ data.  In Sec.~\ref{sec:bcgc} we introduce the impact parameter dependence into the model and determine the optimum value of $\gamma_s$.  The impact parameter dependent saturation scale is discussed in Sec.~\ref{sec:saturation}.  Predictions are obtained for the longitudinal and heavy flavour structure functions in Sec.~\ref{sec:fl} and for exclusive diffractive processes in Sec.~\ref{sec:excl}; these predictions are then confronted with HERA data.  Finally, we conclude in Sec.~\ref{sec:conclusions}.

\section{Colour glass condensate dipole model} \label{sec:cgc}
The CGC dipole cross section of Iancu, Itakura and Munier \cite{Iancu:2003ge}, integrated over impact parameter, can be written as
\begin{equation} \label{eq:cgcfact}
  \sigma_{q\bar q} \equiv \int\!\dif^2\vec{b}\;\frac{\dif\sigma_{q\bar q}}{\dif^2\vec{b}} = 2\int\!\dif^2\vec{b}\;\mathcal{N}(x,r,b) = 2\int\!\dif^2\vec{b}\;T(b)\,\mathcal{N}(x,r) = \sigma_0\,\mathcal{N}(x,r),
\end{equation}
that is, the $b$ dependence of $\mathcal{N}(x,r,b)$, the imaginary part of the dipole--proton scattering amplitude, is assumed to factorise, so that the integration over $\vec{b}$ gives a multiplicative constant $\sigma_0$ determined by a fit to $F_2$ data.  The usual interpretation \cite{Iancu:2003ge} is that the proton is assumed to be a homogeneous disk of radius $R_p$.  Then the $b$ dependence is given by a step function $T(b) = \Theta(R_p-b)$, so that integration over $\vec{b}$ gives $\sigma_0 = 2\pi R_p^2$ in \eqref{eq:cgcfact}.  In fact, $T(b)$ need not necessarily be a step function.  For example, it is pointed out in Ref.~\cite{Marquet:2007nf} that a factorised Gaussian $b$ dependence will also lead to \eqref{eq:cgcfact}.  However, the factorisation of the $b$ dependence from the $x$ dependence is not supported by the HERA diffractive data, where one finds a significantly nonzero effective $\alpha_\Pom^\prime$, indicating correlation between the $b$ and $x$ dependence of the dipole scattering amplitude; see Sec.~\ref{sec:excl}.

The scattering amplitude $\mathcal{N}(x,r)$ can vary between zero and one, where $\mathcal{N}=1$ is the unitarity (``black disc'') limit.  $\mathcal{N}$ is obtained by smoothly interpolating between two limiting types of behaviour.  For small dipole sizes, $r\ll 2/Q_s$, $\mathcal{N}$ is obtained from the saddle point approximation to the leading-order (LO) BFKL equation, followed by an expansion to second order around the saturation saddle point.  For large dipole sizes, $r\gg 2/Q_s$, the functional form obtained from solving the BK equation is used \cite{Levin:1999mw}.  The scattering amplitude is therefore \cite{Iancu:2003ge}
\begin{equation} \label{eq:cgc}
  \mathcal{N}(x,r) =
  \begin{cases}
    \mathcal{N}_0\left(\frac{rQ_s}{2}\right)^{2\left(\gamma_s+\frac{1}{\kappa\lambda Y}\ln\frac{2}{rQ_s}\right)} & :\quad rQ_s\le 2\\
    1-\mathrm{e}^{-A\ln^2(BrQ_s)} & :\quad rQ_s>2
  \end{cases},
\end{equation}
where $Q_s\equiv Q_s(x)=(x_0/x)^{\lambda/2}$ GeV, $Y=\ln(1/x)$, and $\kappa = \chi''(\gamma_s)/\chi'(\gamma_s)$ where $\chi$ is the LO BFKL characteristic function.  The coefficients $A$ and $B$ in the second line of \eqref{eq:cgc} are determined uniquely from the condition that $\mathcal{N}(x,r)$, and its derivative with respect to $rQ_s$, are continuous at $rQ_s=2$:
\begin{equation} \label{eq:AandB}
  A = -\frac{\mathcal{N}_0^2\gamma_s^2}{(1-\mathcal{N}_0)^2\ln(1-\mathcal{N}_0)}, \qquad B = \frac{1}{2}\left(1-\mathcal{N}_0\right)^{-\frac{(1-\mathcal{N}_0)}{\mathcal{N}_0\gamma_s}}.
\end{equation}
For $rQ_s\gtrsim 2$, the scattering amplitude is a function only of $rQ_s$; this is the so-called geometric scaling.  For small dipole sizes, $rQ_s\ll 2$, the second (``diffusion'') term in the exponent of the first line of \eqref{eq:cgc} enhances the effective anomalous dimension\footnote{More precisely, the usual ``anomalous dimension'' is really $1-\gamma_{\rm eff}$, rather than $\gamma_{\rm eff}$ itself \cite{Iancu:2003ge}.} from $\gamma_s$ to
\begin{equation} \label{eq:anomdim}
  \gamma_{\rm eff} \equiv \frac{\partial\ln\mathcal{N}}{\partial\ln\left(r^2 Q_s^2/4\right)} = \gamma_s + \frac{2}{\kappa\lambda Y}\ln\frac{2}{rQ_s}.
\end{equation}
The diffusion term violates geometric scaling, but is essential to describe the data by mimicking the effect of DGLAP evolution.  As $r$ decreases, $\gamma_{\rm eff}$ increases from $\gamma_s$ towards the DGLAP value of 1.  However, $\gamma_{\rm eff}\to \infty$ in the limit that $r\to 0$, in disagreement with the expected colour transparency ($\mathcal{N}\sim r^2$ as $r\to 0$), but this unphysical behaviour has been argued to have no influence on the fit since the contribution to $\sigma^{\gamma^*p}_{\rm tot}$ from very small dipoles is negligible \cite{Iancu:2003ge}.

\begin{table}
  \centering
  \begin{tabular}{cccc|c|c}
    \hline\hline
    $\gamma_s$ & $\sigma_0$/mb & $x_0$ & $\lambda$ & $\chisq$ & $p$-value \\ \hline
    $0.63$ (fixed, \cite{Kowalski:2006hc}) & $35.7$ & $2.70\times10^{-7}$ & $0.177$ & $116.8/130=0.90$ & $0.79$ \\
    $0.74$ (fitted) & $27.4$ & $1.63\times10^{-5}$ & $0.216$ & $110.4/129=0.86$ & $0.88$ \\
    $0.61$ (fitted) & $37.4$ & $1.09\times10^{-7}$ & $0.170$ & $115.4/129=0.89$ & $0.80$ \\
    $0.44$ (fixed) & $46.3$ & $2.21\times10^{-11}$ & $0.122$ & $180.1/130=1.39$ & $2.4\times10^{-3}$ \\
    \hline\hline
  \end{tabular}
  \caption{Parameters of the CGC dipole model \eqref{eq:cgc}, for different values of $\gamma_s$, determined from fits to ZEUS $F_2$ data \cite{Breitweg:2000yn,Chekanov:2001qu} with $\xB\le0.01$ and $Q^2\in[0.25,45]$ GeV$^2$.  The meaning of the $p$-value is explained in the text.}
  \label{tab:cgc}
\end{table}

The original CGC fit \cite{Iancu:2003ge} (and the subsequent fit including charm quarks \cite{Kowalski:2006hc}) fixed the parameters $\gamma_s=0.63$ and $\kappa=9.9$ at the LO BFKL values.  The central fits were obtained with $\mathcal{N}_0$ fixed at $0.7$ and the other three parameters ($\sigma_0$, $x_0$ and $\lambda$) were fitted to $F_2$ data \cite{Breitweg:2000yn,Chekanov:2001qu} with $\xB\le0.01$ and $Q^2\le45$ GeV$^2$.  The results from the fit including charm quarks \cite{Kowalski:2006hc} are shown in the first line of Table \ref{tab:cgc}.  Statistical and systematic experimental errors are added in quadrature.  The quark masses in the photon wave functions were taken to be $m_{u,d,s}=0.14$ GeV and $m_c=1.4$ GeV, and the dipole cross section was evaluated at $x=\xB$ for light quarks and $x=\xB(1+4m_c^2/Q^2)$ for charm quarks.  The contribution from beauty quarks was neglected.  The saturation scale was found to drop dramatically with the introduction of charm; for example, the $x_0$ parameter was lowered by two orders of magnitude \cite{Kowalski:2006hc}.  However, it has recently been shown by Soyez \cite{Soyez:2007kg} that allowing $\gamma_s$ to vary from the LO BFKL value of $0.63$ gives an improved fit to the $F_2$ data with a $\gamma_s$ slightly larger than $0.7$, close to the value extracted from renormalisation-group-improved next-to-leading order (NLO) BFKL kernels.  Moreover, the saturation scale is not significantly reduced compared to the saturation scale obtained with only light quarks.  The results of this fit \cite{Soyez:2007kg} have already been applied to describe inclusive diffractive DIS \cite{Marquet:2007nf} and exclusive diffractive processes using a $t$-dependent saturation scale \cite{Marquet:2007qa}.

In the second line of Table \ref{tab:cgc} we show the results of a fit to $F_2$ data starting from the parameters favoured by Soyez \cite{Soyez:2007kg} (keeping $\kappa$ fixed at 9.9).  Indeed, the $\chi^2$ is slightly improved compared to the first fit with a fixed $\gamma_s=0.63$.  The results are not identical to those of Soyez \cite{Soyez:2007kg} due to our lack of beauty quark contribution, the fact that we do not include H1 data, and our more conservative $Q^2$ cuts, but none of these differences has much effect on the parameters obtained.  (As in earlier analyses \cite{Iancu:2003ge,Kowalski:2003hm,Forshaw:2004vv,Kowalski:2006hc} we do not fit the H1 data to avoid introducing an extra normalisation parameter and because the ZEUS data alone suffice.\footnote{However, new preliminary low-$Q^2$ inclusive DIS data from H1 \cite{VargasTrevino:2007zz} will provide tighter constraints on future dipole model fits.}  We do not include data with $Q^2>45$ GeV$^2$ since the CGC model does not include the full DGLAP evolution, which becomes more important at large $Q^2$.  We do not include data with $Q^2<0.25$ GeV$^2$ because the form of the CGC dipole cross section is motivated by perturbative QCD, although including these few data points has little effect on the parameters obtained.)

In the third line of Table \ref{tab:cgc} we show the parameters of a local minimum which is close to the first fit \cite{Kowalski:2006hc}, but has a slightly smaller $\gamma_s = 0.61$.  Using a hypothesis-testing criterion \cite{Collins:2001es}, a ``good'' fit with $N$ degrees of freedom should have a $\chi^2$ of approximately $N\pm\sqrt{2N}$.  More precisely, we can calculate the $p$-value of the hypothesis shown in the last column of Table \ref{tab:cgc}, defined as the probability, under the assumption of a given hypothesis, of obtaining data at least as incompatible with the hypothesis as the data actually observed \cite{Yao:2006px}.  Then, for example, the hypothesis could be considered to be excluded at a 90\% confidence level if $1-p>0.9$.  The $\chi^2$ of the first three fits in Table \ref{tab:cgc} can all be considered to be ``good'' according to a hypothesis-testing criterion, therefore it is difficult to say that the fit with $\gamma_s = 0.74$ is strongly preferred by the data compared to the fit with $\gamma_s = 0.61$.

Recent studies of the numerical solution of the BK equation for a fixed impact parameter have shown that the effective anomalous dimension $\gamma_{\rm eff}$ is not a constant at the saturation scale, but approaches a limiting value of $\gamma_s \simeq 0.44$ in the small-$x$ limit \cite{Boer:2007wf}.\footnote{In Ref.~\cite{Boer:2007wf} the LO BK equation with a fixed coupling $\bar{\alpha}_S = 0.2$ was used, which gives a very large value of $\lambda \simeq 0.9$.  It was argued that the introduction of a running coupling, giving a smaller value of $\lambda$, was questionable given that the BK equation is LO in $\alpha_S$, and that the running coupling did not have a significant effect on the results obtained for the anomalous dimension.}  The value of $\gamma_s \simeq 0.44$ \cite{Boer:2007wf} seems to be inconsistent with the phenomenological values found in the CGC model.  Fixing $\gamma_s = 0.44$ and allowing the other parameters to go free gives the results shown in the fourth line of Table \ref{tab:cgc}; this cannot be considered a good fit.  It is interesting to check if the same conclusions are found in the more realistic impact parameter dependent version of the CGC model.

\section{Introducing the impact parameter dependence} \label{sec:bcgc}
To introduce the impact parameter dependence into the CGC model \cite{Iancu:2003ge}, we modify \eqref{eq:cgc} to obtain the ``b-CGC'' model for the dipole cross section \cite{Iancu,Kowalski:2006hc}:
\begin{equation} \label{eq:bcgc}
  \frac{\dif\sigma_{q\bar{q}}}{\dif^2\vec{b}} = 2\,\mathcal{N}(x,r,b)=2\times\begin{cases}
  \mathcal{N}_0\left(\frac{rQ_s}{2}\right)^{2\left(\gamma_s+\frac{1}{\kappa\lambda Y}\ln\frac{2}{rQ_s}\right)} & :\quad rQ_s\le 2\\
  1-\mathrm{e}^{-A\ln^2(BrQ_s)} & :\quad rQ_s>2
  \end{cases},
\end{equation}
where the coefficients $A$ and $B$ in the second line are again given by \eqref{eq:AandB}.  The saturation scale $Q_s$ now depends on the impact parameter:
\begin{equation} \label{eq:bcgc1}
  Q_s\equiv Q_s(x,b)=\left(\frac{x_0}{x}\right)^{\frac{\lambda}{2}}\;\left[\exp\left(-\frac{b^2}{2B_{\rm CGC}}\right)\right]^{\frac{1}{2\gamma_s}}.
\end{equation}
Instead of the normalisation parameter $\sigma_0$ of the CGC model, we now have the parameter $B_{\rm CGC}$, which is adjusted iteratively to give a good description of the $t$ dependence of exclusive $J/\psi$ photoproduction; see Sec.~\ref{sec:excl}.  We also allow the factor $\mathcal{N}_0$ to go free.  The fit presented in Ref.~\cite{Kowalski:2006hc} fixed $\gamma_s = 0.63$, but was unsuccessful in obtaining a good fit, as seen in the first line of Table \ref{tab:bcgc}.  For this reason, Ref.~\cite{Kowalski:2006hc} mostly concentrated on the b-Sat model description of exclusive processes, that is, using the Glauber--Mueller dipole cross section with DGLAP evolution of the gluon density.  Nevertheless, the b-CGC fit presented in Ref.~\cite{Kowalski:2006hc} has already been applied to calculate electroweak deeply virtual Compton scattering \cite{Machado:2007wq}, quarkonium photoproduction in coherent hadron--hadron interactions \cite{Goncalves:2007sa}, the saturation scale in large nuclei \cite{Kowalski:2007rw}, and diffractive structure functions for both protons and nuclei \cite{Kowalski:2008sa}.

\begin{table}
  \centering
  \begin{tabular}{ccccc|c|c}
    \hline\hline
    $\gamma_s$ & $B_{\rm CGC}$/GeV$^{-2}$ & $\mathcal{N}_0$ & $x_0$ & $\lambda$ & $\chisq$ & $p$-value \\ \hline
    $0.63$ (fixed, \cite{Kowalski:2006hc}) & $5.5$ & $0.417$ & $5.95\times10^{-4}$ & $0.159$ & $211.2/130=1.62$ & $8.7\times 10^{-6}$ \\ 
    $\mathbf{0.46}$ & $\mathbf{7.5}$ & $\mathbf{0.558}$ & $\mathbf{1.84\times10^{-6}}$ & $\mathbf{0.119}$ & $\mathbf{118.7/129=0.92}$ & $\mathbf{0.73}$ \\
    $0.43$ (no sat.) & $7.5$ & $0.565$ & $1.34\times10^{-6}$ & $0.109$ & $124.3/129=0.96$ & $0.60$ \\
    $0.54$ (high-$Q^2$) & $6.5$ & $0.484$ & $3.42\times10^{-5}$ & $0.149$ & $210.9/159=1.33$ & $3.7\times 10^{-3}$ \\
    \hline\hline
  \end{tabular}
  \caption{Parameters of the b-CGC model, \eqref{eq:bcgc} and \eqref{eq:bcgc1}, determined from fits to ZEUS $F_2$ data \cite{Breitweg:2000yn,Chekanov:2001qu} with $\xB\le0.01$ and $Q^2\in[0.25,45]$ GeV$^2$.  The first line is the fit \cite{Kowalski:2006hc} with a fixed $\gamma_s=0.63$, while the second line is the new fit where $\gamma_s$ is allowed to go free.  The third line is a fit without explicit saturation of the dipole cross section, that is, the form of \eqref{eq:bcgc} for $rQ_s\le 2$ is also taken for $rQ_s>2$ (but without the diffusion term in the exponent).  The fourth line is a fit which also includes high-$Q^2$ DIS data up to $Q^2\le650$ GeV$^2$.}
  \label{tab:bcgc}
\end{table}

We find that allowing $\gamma_s$ to vary in addition to the other parameters dramatically improves the description of the $F_2$ data, as seen in the second line of Table \ref{tab:bcgc}.  Moreover, the optimum value of $\gamma_s = 0.46$ is close to the expected value of $0.44$ obtained from the numerical solution of the BK equation \cite{Boer:2007wf}.  The fact that the optimum value of $\gamma_s$ is quite different in the CGC and b-CGC models, and that the value in the b-CGC model is closer to the theoretical expectation, may be attributed to the more realistic modelling of the impact parameter dependence in the b-CGC model compared to the factorised $b$ dependence implicit in the CGC model.  However, the value of $\lambda = 0.119$ obtained from the fit is lower than the perturbatively calculated value of $\lambda\sim 0.3$ \cite{Triantafyllopoulos:2002nz}, and suggests that the saturation scale comprises significant nonperturbative dynamics.

To examine the importance, or otherwise, of having a unitarised dipole cross section, we performed a fit without explicit saturation of the dipole cross section, that is, the form of \eqref{eq:bcgc} for $rQ_s\le 2$ is also taken for $rQ_s>2$ (but without the diffusion term in the exponent).  The results of this fit are shown in the third line of Table \ref{tab:bcgc}.  The fit is only slightly worse, and the parameters only slightly different, than the main fit presented in the second line of Table \ref{tab:bcgc}.  This suggests that the HERA data fitted are largely insensitive to the presence of saturation.

Taking the parameters given in the second line of Table \ref{tab:bcgc}, and releasing the upper cut on $Q^2$ gives a $\chi^2$ of 381 for the 163 data points.  For comparison, the ``b-Sat'' model gave a $\chi^2$ of 193 for the same data \cite{Kowalski:2006hc}.  Performing a fit improves the $\chi^2$ to 211 with parameters given in the fourth line of Table \ref{tab:bcgc}.  Since the b-CGC model is not justified for large $Q^2$, we favour the fit with the more conservative cut of $Q^2\le 45$ GeV$^2$, that is, with parameters given in the second line of Table \ref{tab:bcgc}.

\section{Impact parameter dependent saturation scale} \label{sec:saturation}
It is customary to define a saturation scale $Q_S$, that is, the momentum scale at which the dipole--proton scattering amplitude $\mathcal{N}$ becomes sizable such that nonlinear effects start to become important.  There is no unique definition of $Q_S$ and various choices are used in the literature.  Following Refs.~\cite{Kowalski:2003hm,Kowalski:2006hc} we define the saturation scale $Q_S^2\equiv 2/r_S^2$, where the saturation radius $r_S$ is the dipole size where the scattering amplitude
\begin{equation} \label{eq:satdef}
  \mathcal{N} = 1 - \mathrm{e}^{-\frac{1}{2}} \simeq 0.4.
\end{equation}
The saturation scale $Q_S^2=2/r_S^2$ defined by \eqref{eq:satdef} coincides with $Q_s^2(x) \equiv 1/R_0^2(x) = (x_0/x)^{\lambda}$ GeV$^2$ in the GBW model \cite{Golec-Biernat:1998js,Golec-Biernat:1999qd}.  However, for the CGC \eqref{eq:cgc} and b-CGC \eqref{eq:bcgc} models, the saturation scale $Q_S$ defined by \eqref{eq:satdef} differs from the parameter $Q_s$.  Note that we use uppercase $S$ and lowercase $s$ to distinguish between these two scales.  The saturation scale $Q_S$ is the quantity we shall compute and compare for the different dipole models.

\begin{figure}
  \centering
  \includegraphics[width=0.8\textwidth]{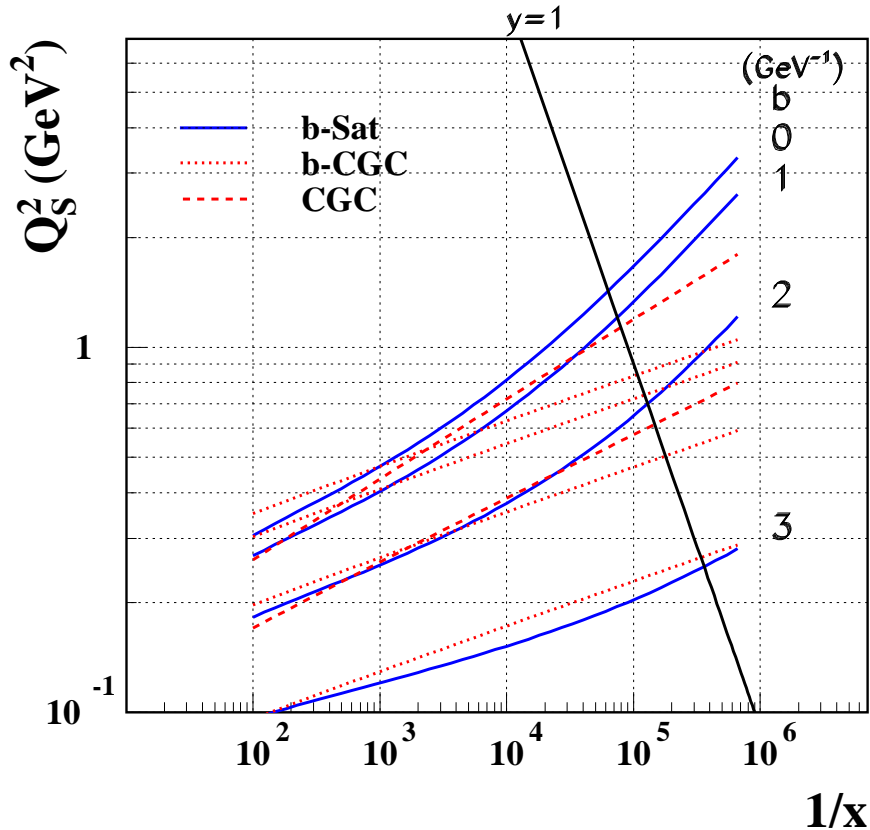}
  \caption{The impact parameter dependent saturation scale $Q_S^2\equiv 2/r_S^2$, where $r_S$ is defined as the solution of \eqref{eq:satdef}, at $b=0,1,2,3$ GeV$^{-1}$, found in the b-Sat \cite{Kowalski:2006hc} and b-CGC models (second line of Table \ref{tab:bcgc}).  We also show the saturation scale from the two candidate solutions of the $b$-independent CGC model in Table \ref{tab:cgc} with $\gamma_s = 0.74$ (upper dashed line) and $\gamma_s = 0.61$ (lower dashed line).  We indicate the HERA kinematic limit at $y\simeq Q^2/(\xB\,s)=1$ with $s\simeq 4E_eE_p=90200$ GeV$^2$.}
    \label{fig:q2sx}
\end{figure}

In Fig.~\ref{fig:q2sx} we show the impact parameter dependent saturation scale $Q_S^2$ for the b-Sat model \cite{Kowalski:2006hc} (solid lines) and the present b-CGC model (dotted lines) at $b=0,1,2,3$ GeV$^{-1}$.  The saturation scale $Q_S^2$ is strongly dependent on the impact parameter $b$.  We also show $Q_S^2$ for the two candidate solutions for the $b$-independent CGC model (dashed lines) with $\gamma_s = 0.74$ and $\gamma_s = 0.61$.

\begin{figure}
  \centering
  \includegraphics[width=0.5\textwidth]{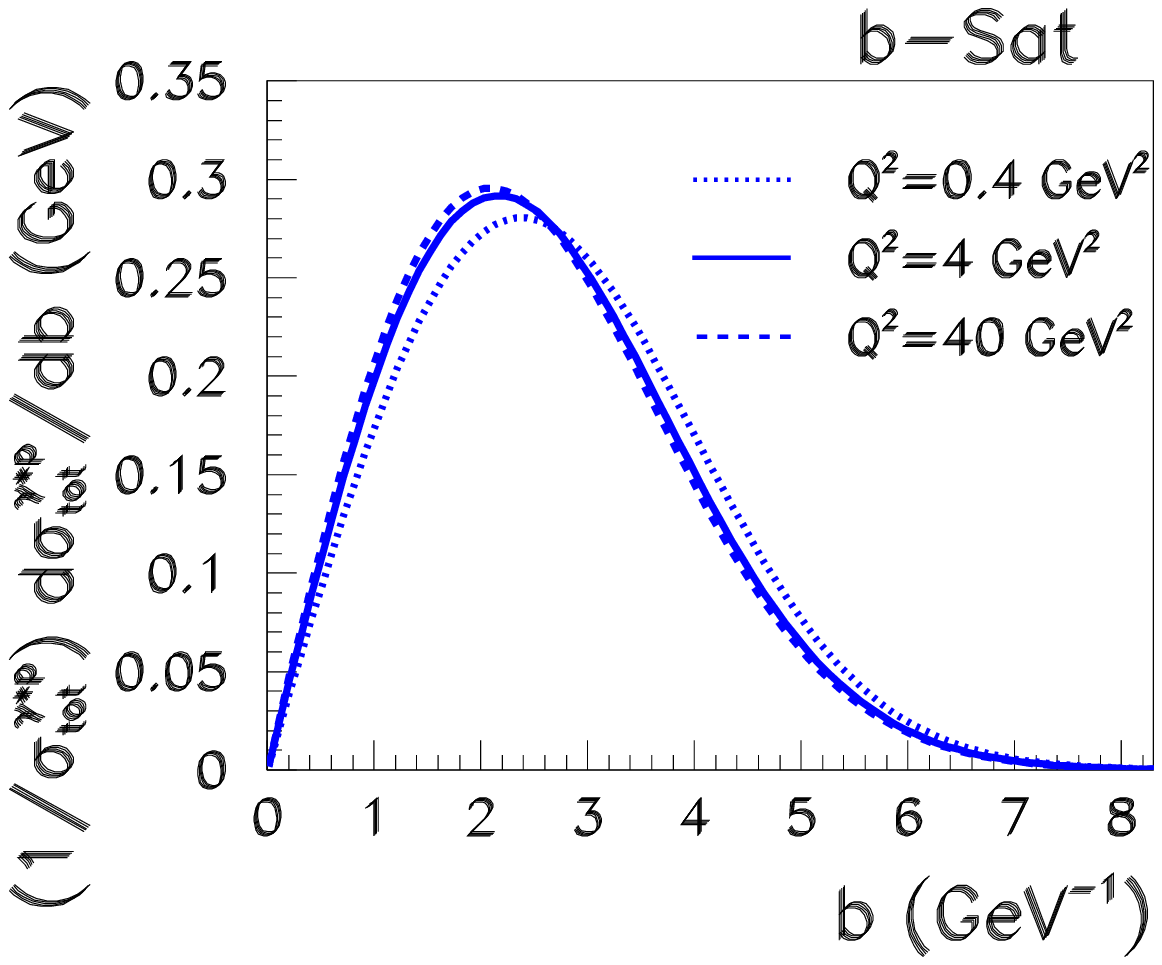}%
  \includegraphics[width=0.5\textwidth]{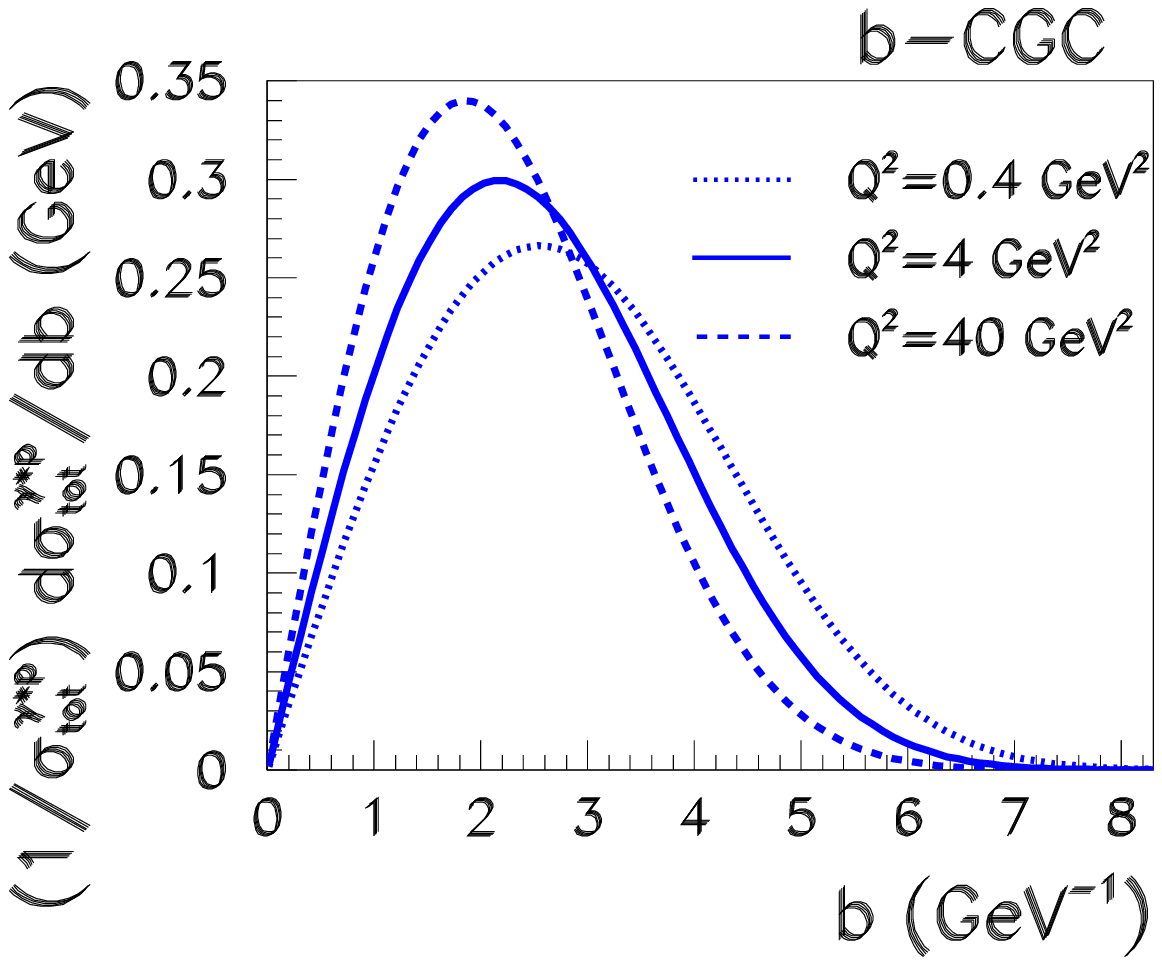}
  \caption{The $b$ dependence of the total $\gamma^*p$ cross section, $\sigma_{\rm tot}^{\gamma^*p}$, for $Q^2=0.4$, $4$ and $40$ GeV$^2$ with $x=10^{-4}$, $10^{-3}$ and $10^{-2}$ respectively, in the b-Sat model (left) and the b-CGC model (right).  The median values of $b$ are all between $2$ and $3$ GeV$^{-1}$.}
  \label{fig:psitlb}
\end{figure}

In Fig.~\ref{fig:psitlb} we show the $b$ dependence in the b-Sat and b-CGC models of the total $\gamma^*p$ cross section for representative values of $x$ and $Q^2$ of the HERA $F_2$ data included in the fits.  The $b$ dependence is approximately factorised from the other kinematic variables in the b-Sat model \cite{Kowalski:2006hc}, unlike the case in the b-CGC model.  For both models the median values of $b$ probed in the total $\gamma^*p$ cross section are in the range $2$--$3$ GeV$^{-1}$.  Therefore, from Fig.~\ref{fig:q2sx}, the solution of the $b$-independent CGC model with $\gamma_s = 0.61$ has a similar $Q_S^2$ to the b-CGC model at some average value of $b$, and this is much lower than the $Q_S^2$ of the solution with $\gamma_s = 0.74$ favoured by Soyez \cite{Soyez:2007kg}.
Although the b-Sat and b-CGC parameterisations of the dipole cross section have very different theoretical motivations, both give a similar saturation scale $Q_S^2\lesssim 0.5$ GeV$^2$ for $x\gtrsim 10^{-5}$ for the most relevant impact parameters $b\sim 2$--$3$ GeV$^{-1}$.

The saturation aspect of the dipole models has been emphasised since the initial GBW investigations \cite{Golec-Biernat:1998js,Golec-Biernat:1999qd}.  However, as the dipole models have become more sophisticated, with the introduction of DGLAP evolution, heavy quarks, and impact parameter dependence, the saturation scale has lowered and saturation is no longer crucial in describing the HERA data.  Two-component Regge-motivated dipole models (for example, Ref.~\cite{Forshaw:2004vv}) find some preference for saturation, but these do not include DGLAP evolution or impact parameter dependence.  Indeed, temporary modifications of the b-Sat and b-CGC models, so that the dipole cross sections no longer unitarise at large dipole sizes and so only the single-Pomeron exchange is present, give comparable fits with only a slight change in the parameters required, as seen for the b-Sat model in Ref.~\cite{Kowalski:2003hm} and for the b-CGC model in the third line of Table \ref{tab:bcgc}.  Therefore, the fact that a model incorporating saturation is successful in describing the data should \emph{not} be construed as meaning that there are large saturation effects present in the data.

Of course, from the theoretical point-of-view, the dipole model is only self-consistent if $S$-matrix unitarity is imposed, that is, the scattering amplitude $\mathcal{N}$ cannot take on values greater than one.  This feature is necessary to derive relations between the inclusive cross section and cross sections for diffractive processes.  Moreover, dipole models incorporating saturation fitted to HERA data may be extrapolated to very low $x$ (for example, at the LHeC \cite{Dainton:2006wd}) and to predict cross sections for nuclear collisions where the saturation scale is enhanced by $A^{1/3}$ \cite{Kowalski:2007rw} (for example, at a future electron--ion collider \cite{Deshpande:2005wd}).  In these situations, multi-Pomeron exchange may become important and extrapolation based on single-Pomeron exchange would be unreliable.

\section{Description of the longitudinal and heavy flavour structure functions} \label{sec:fl}

We now compare the predictions of the present b-CGC model, and also those from the b-Sat model \cite{Kowalski:2006hc}, with published HERA data on the longitudinal structure function, $F_L(x,Q^2)$, the charm structure function, $F_2^{c\bar{c}}(x,Q^2)$, and the beauty structure function, $F_2^{b\bar{b}}(x,Q^2)$.  These predictions can easily be obtained by taking the appropriate contributions to the total $\gamma^*p$ cross section \eqref{eq:siggp}.

\begin{figure}
  \includegraphics[width=\textwidth]{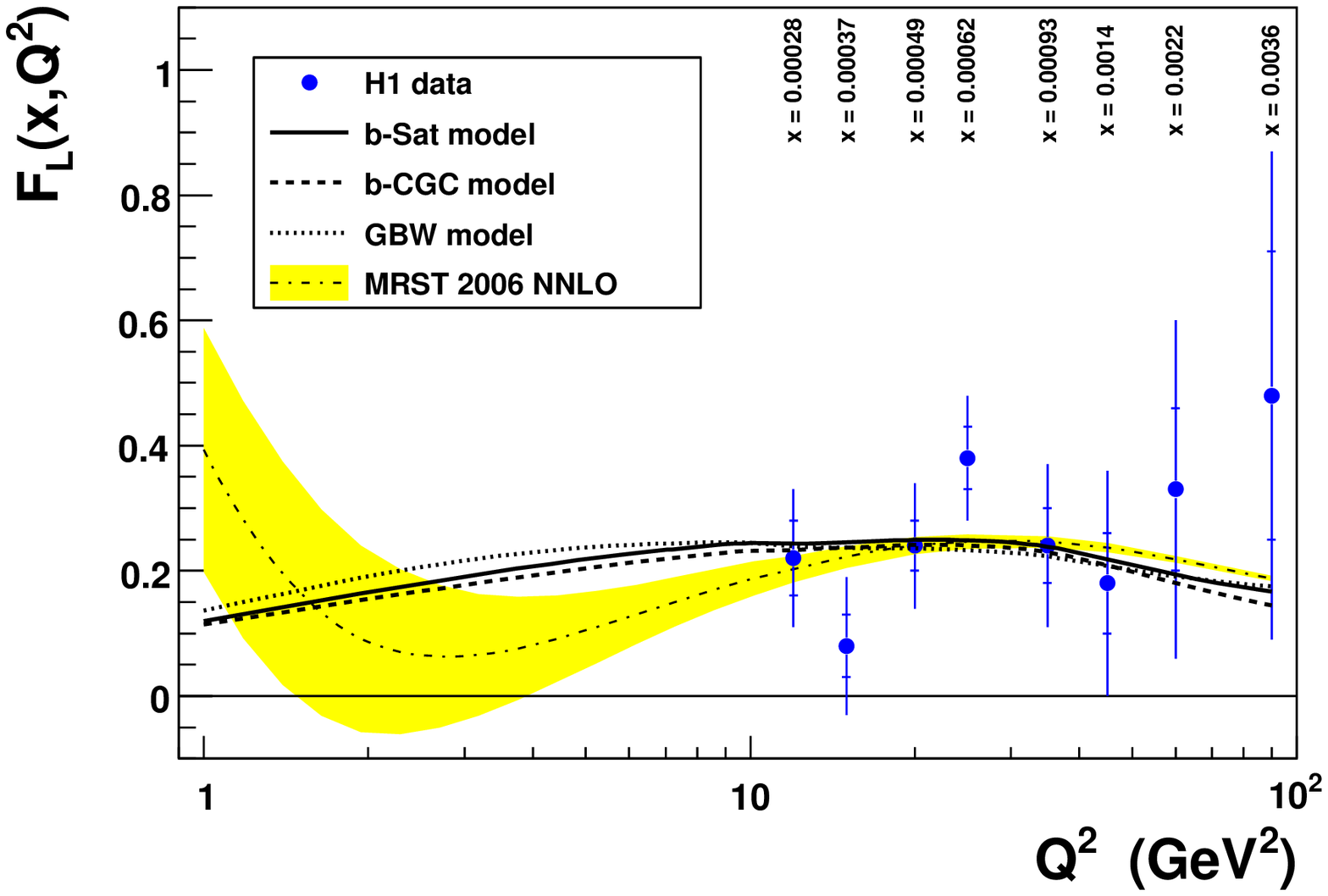}
  \caption{Predictions for the longitudinal structure function, $F_L(x,Q^2)$, using three different dipole model cross sections compared to H1 data \cite{Aaron:2008tx}.  We also show predictions calculated at NNLO in the collinear factorisation framework using the MRST 2006 NNLO parton distributions \cite{Martin:2007bv}.}
  \label{fig:h1fl}
\end{figure}
A first measurement has been reported \cite{Aaron:2008tx} of the longitudinal structure function $F_L(x,Q^2)$ for $Q^2\in [12,90]$ GeV$^2$ based on data taken in the last few months of HERA running, when the proton beam energy was lowered from the nominal value of 920 GeV to values of 460 GeV and 575 GeV.  In Fig.~\ref{fig:h1fl} we show these data \cite{Aaron:2008tx} compared to predictions calculated using \eqref{eq:siggp} with three different dipole cross sections: the b-Sat model \cite{Kowalski:2006hc}, the present b-CGC model, and the GBW fit including charm quarks from Ref.~\cite{Kowalski:2006hc}.  We also show the predictions calculated at next-to-next-to-leading order (NNLO) in the conventional collinear factorisation framework using the MRST 2006 NNLO parton distributions \cite{Martin:2007bv}, including the uncertainty band obtained from the 30 alternative eigenvector parton distribution sets.

In the region of the existing H1 data \cite{Aaron:2008tx} the three dipole model predictions all agree well with the data and also with the NNLO calculation.  However, we also show in Fig.~\ref{fig:h1fl} the predictions at lower $Q^2\lesssim 10$ GeV$^2$.  We extrapolate the values of $x$ accordingly, by fitting the $x$ values of the H1 data as a power law function of $Q^2$, obtaining $x = (1.09\times10^{-5})(Q^2/Q_0^2)^{1.28}$ with $Q_0^2 = 1$ GeV$^2$.  Below 10 GeV$^2$ the dipole model predictions deviate from the NNLO predictions and lie well outside the uncertainty band.  There is little sensitivity to the detailed form of the dipole cross section, as shown by the consistency between the three predictions.  The dipole model calculations contain some effects from small-$x$ resummation and higher-twist contributions that are not included in the fixed-order collinear factorisation approach, which is known to be perturbatively unstable at low $x$ and $Q^2$.  A measurement of $F_L$ at low $x$ and $Q^2$ would therefore play an important r\^ole in discriminating between the different theoretical approaches.  If not at HERA, such a measurement could be possible at the LHeC \cite{Dainton:2006wd} or at a future electron--ion collider \cite{Deshpande:2005wd}.

\begin{figure}
  \includegraphics[width=0.9\textwidth]{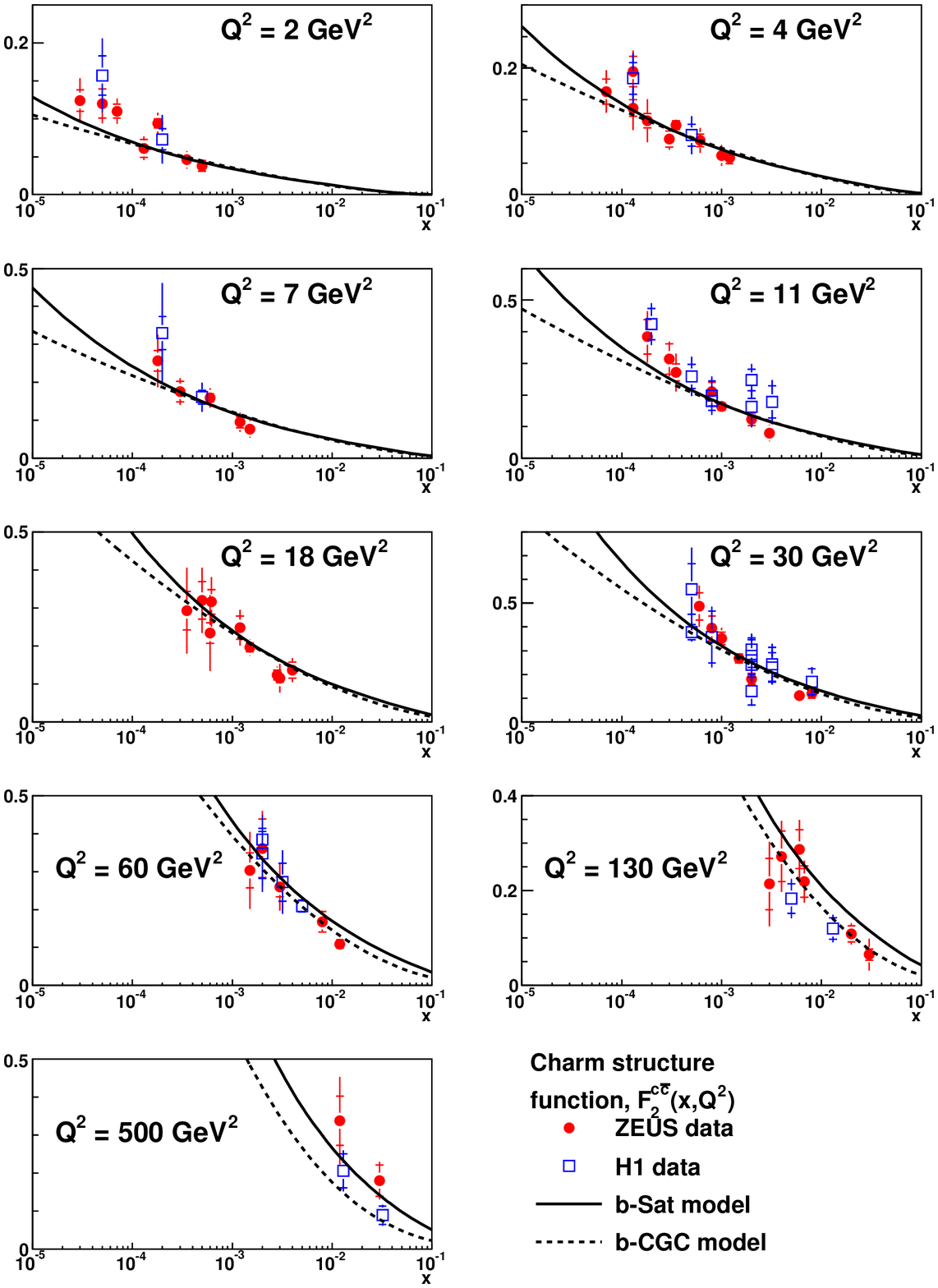}
  \caption{Predictions for the charm structure function, $F_2^{c\bar{c}}(x,Q^2)$, using two different dipole model cross sections compared to data from ZEUS \cite{Breitweg:1999ad,Chekanov:2003rb,Chekanov:2007ch} and H1 \cite{Adloff:1996xq,Adloff:2001zj,Aktas:2005iw,Aktas:2004az}.  The $Q^2$ bins used correspond to the data points in Ref.~\cite{Chekanov:2003rb}; the other data points have been shifted to these $Q^2$ values using the b-Sat model predictions.}
  \label{fig:f2charm}
\end{figure}
In Fig.~\ref{fig:f2charm} we show the b-Sat and b-CGC predictions for the charm structure function $F_2^{c\bar{c}}(x,Q^2)$ compared to all available ZEUS \cite{Breitweg:1999ad,Chekanov:2003rb,Chekanov:2007ch} and H1 \cite{Adloff:1996xq,Adloff:2001zj,Aktas:2005iw,Aktas:2004az} data.  Both models give an equally good description of the data, even for large $x\gtrsim 0.01$ and large $Q^2$ where the b-CGC model might be expected to fail.

\begin{figure}
  \includegraphics[width=0.9\textwidth]{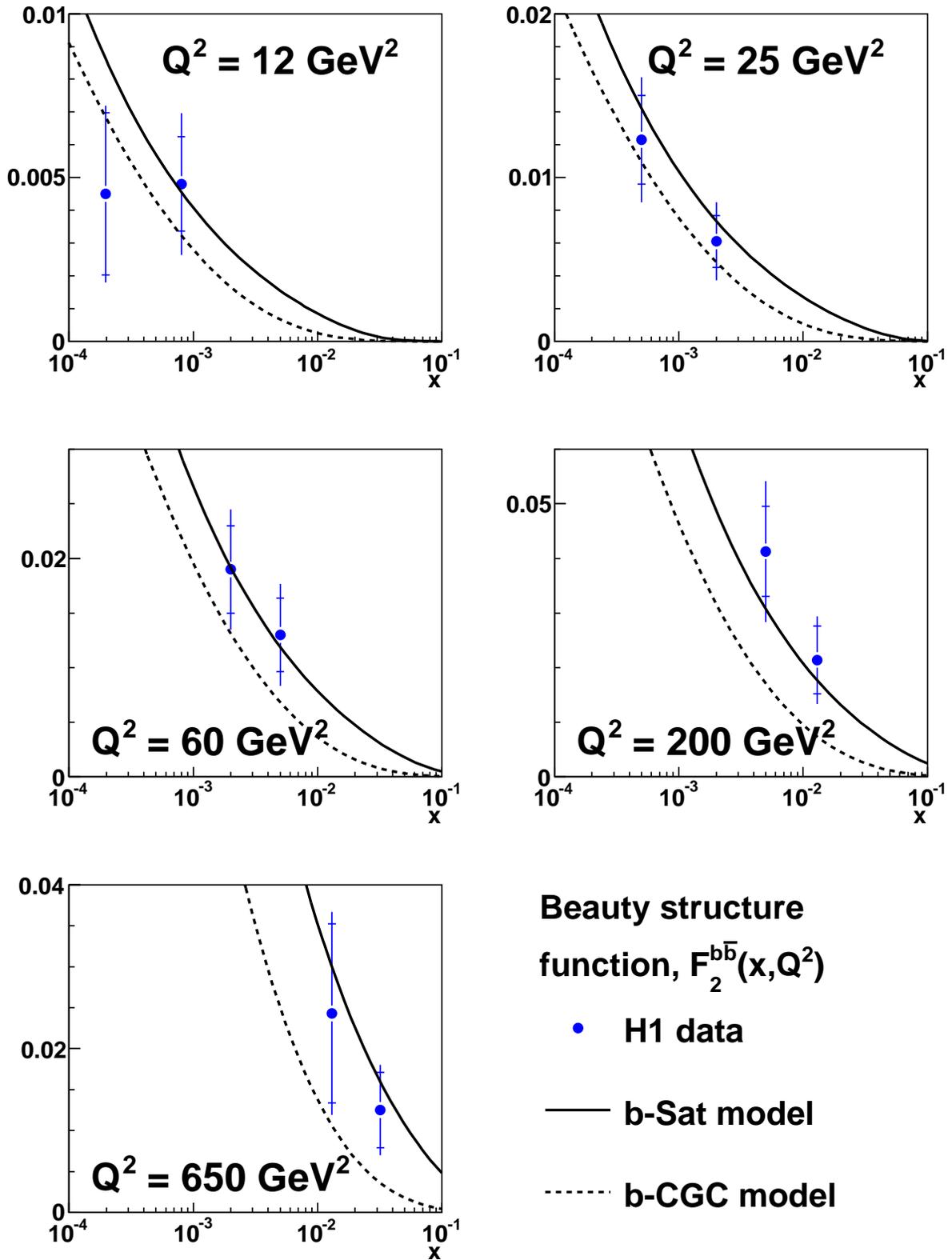}
  \caption{Predictions for the beauty structure function, $F_2^{b\bar{b}}(x,Q^2)$, using two different dipole model cross sections compared to data from H1 \cite{Aktas:2005iw,Aktas:2004az}.}
  \label{fig:f2beauty}
\end{figure}
In Fig.~\ref{fig:f2beauty} we show the predictions for the beauty structure function $F_2^{b\bar{b}}(x,Q^2)$ using a bottom quark mass of $m_b = 4.5$ GeV.  (There is some sensitivity to this value, especially at low $Q^2$.)  Although the available data have large statistical uncertainties, the difference between the b-Sat and b-CGC model predictions increases as $Q^2$ increases, such that the b-Sat model predictions are favoured over those from the b-CGC model in the higher $Q^2$ bins.

\section{Description of exclusive diffractive processes at HERA} \label{sec:excl}
\begin{figure}
  \hspace{0.2\textwidth}b-Sat model\hspace{0.4\textwidth}b-CGC model\hspace*{0.1\textwidth}\ \\
  \begin{minipage}{0.5\textwidth}
    \centering
    \includegraphics[height=4.5cm]{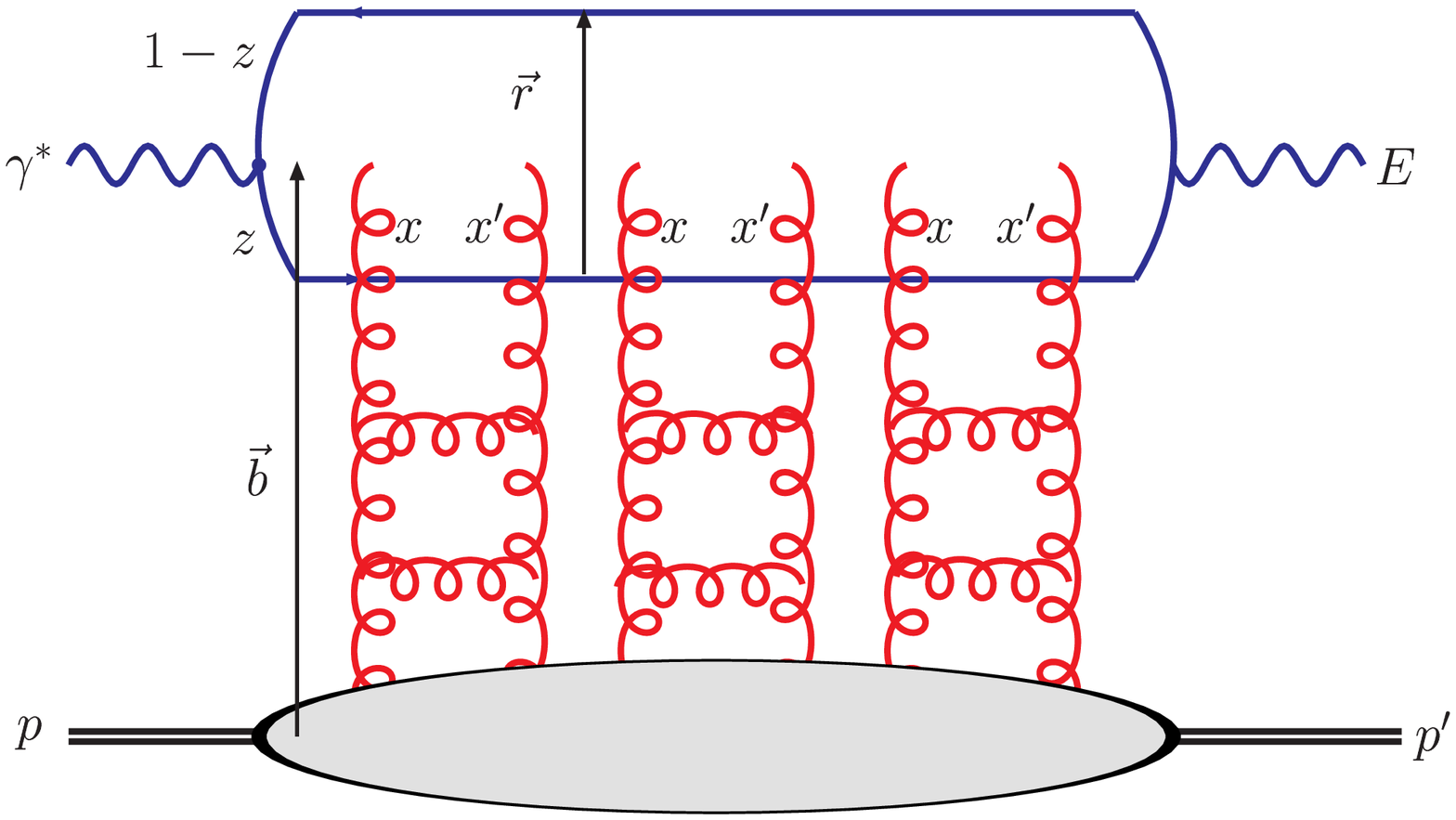}
  \end{minipage}%
  \begin{minipage}{0.5\textwidth}
    \centering
    \includegraphics[height=4.5cm]{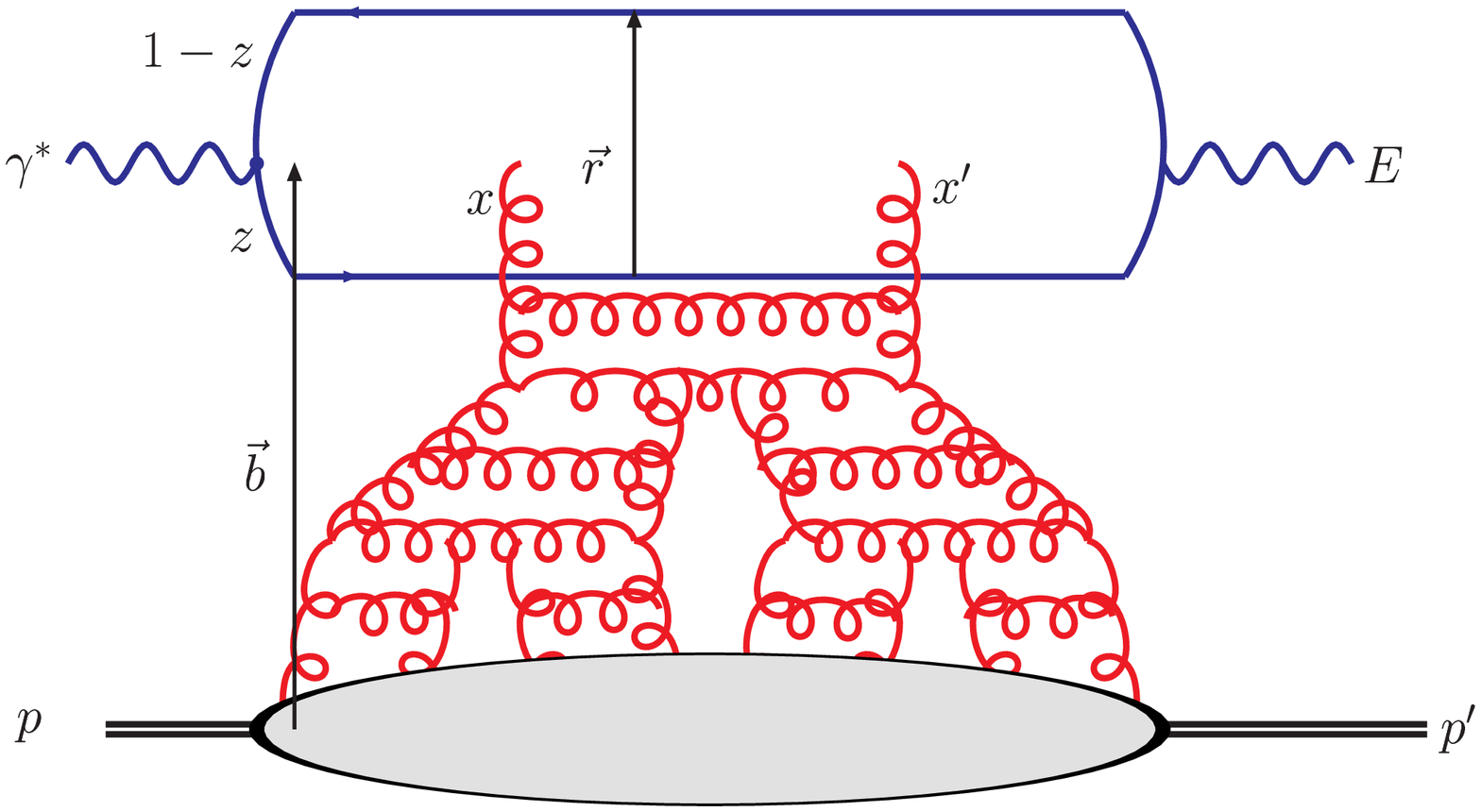}
  \end{minipage}
  \caption{The $\gamma^*p$ scattering amplitude in the b-Sat model (left) and the b-CGC model (right).  For inclusive DIS, we have $E=\gamma^*$, $x=x^\prime\ll 1$ and $p=p^\prime$.  For exclusive diffractive processes, such as vector meson production ($E=V=J/\psi,\phi,\rho$) or DVCS ($E=\gamma$), we have $x^\prime\ll x\ll 1$ and $t=(p-p^\prime)^2$.  The dipole cross section is unitarised assuming eikonalisation of the $b$-dependent DGLAP-evolved gluon density in the b-Sat model (left), or via the fan diagrams resummed by the BK equation in the b-CGC model (right).}
  \label{fig:diag}
\end{figure}
We now check that the updated b-CGC model, with the parameters given in the second line of Table \ref{tab:bcgc}, describes the main features of exclusive diffractive vector meson ($J/\psi$ \cite{Chekanov:2002xi,Chekanov:2004mw,Aktas:2005xu}, $\phi$ \cite{Chekanov:2005cq} and $\rho$ \cite{Adloff:1999kg,Chekanov:2007zr}) production and deeply virtual Compton scattering (DVCS) \cite{Aktas:2005ty,Chekanov:2003ya,Aaron:2007cz,ZEUSDVCS} at HERA, and compare the predictions with the b-Sat model \cite{Kowalski:2006hc}.  The main concepts of the b-Sat and b-CGC models are illustrated in Fig.~\ref{fig:diag}.  The new precise data on $\rho$ electroproduction from ZEUS \cite{Chekanov:2007zr} and new data on DVCS from H1 \cite{Aaron:2007cz} and ZEUS \cite{ZEUSDVCS}, which were not available at the time of Ref.~\cite{Kowalski:2006hc}, provide tighter constraints on the models.  The data have been measured as a function of the photon virtuality, $Q^2$, the $\gamma^*p$ centre-of-mass energy, $W$, and the four-momentum transfer at the proton vertex, $t$.  In addition, the ratio of the cross sections for longitudinally and transversely polarised photons has been measured.  It is therefore a significant challenge for an essentially parameter-free model to describe all features of the available data.

The formalism for calculation of exclusive diffractive processes from the $b$-dependent dipole cross section was discussed in detail in Ref.~\cite{Kowalski:2006hc}.  Here we give only the final formulae.  The differential cross section for the exclusive process $\gamma^*p\to Ep$, where $E=V,\gamma$, is \cite{Kowalski:2006hc}
\begin{equation}
  \frac{\dif\sigma^{\gamma^* p\rightarrow Ep}_{T,L}}{\dif t} = \frac{1}{16\pi}\left\lvert\mathcal{A}^{\gamma^* p\rightarrow Ep}_{T,L}\right\rvert^2\;(1+\beta^2)\,R_g^2,
  \label{eq:xvecm1}
\end{equation}
where the scattering amplitude
\begin{equation} \label{eq:exclamp}
  \mathcal{A}^{\gamma^* p\rightarrow Ep}_{T,L} = \mathrm{i}\,\int\!\dif^2\vec{r}\int_0^1\!\frac{\dif{z}}{4\pi}\int\!\dif^2\vec{b}\;(\Psi_{E}^{*}\Psi)_{T,L}\;\mathrm{e}^{-\mathrm{i}[\vec{b}-(1-z)\vec{r}]\cdot\vec{\Delta}}\;\frac {\dif\sigma_{q\bar q}}{\dif^2\vec{b}}.
\end{equation}
For DVCS, the amplitude involves a sum over quark flavours $f=u,d,s,c$.  The ratio of the real to imaginary parts of the scattering amplitude, $\beta$, is calculated using
\begin{equation} \label{eq:beta}
  \beta = \tan\left(\frac{\pi\lambda}{2}\right), \quad\text{with}\quad \lambda \equiv \frac{\partial\ln\left(\mathcal{A}_{T,L}^{\gamma^* p\rightarrow Ep}\right)}{\partial\ln(1/x)}.
\end{equation}
The factor $R_g^2$ in \eqref{eq:xvecm1} accounts for the skewedness effect, that is, $x\ne x^\prime$ in Fig.~\ref{fig:diag}, and is calculated using \cite{Shuvaev:1999ce}
\begin{equation} \label{eq:Rg}
  R_g(\lambda) = \frac{2^{2\lambda+3}}{\sqrt{\pi}}\frac{\Gamma(\lambda+5/2)}{\Gamma(\lambda+4)},
\end{equation}
with $\lambda$ calculated as in \eqref{eq:beta}.\footnote{In Ref.~\cite{Kowalski:2006hc} the factor $R_g^2$ in \eqref{eq:xvecm1} was omitted for the b-CGC model.  For the b-Sat model, the factor $R_g$ instead multiplies the gluon density inside the dipole cross section.}
The forward overlap functions between the photon and exclusive final state wave functions in \eqref{eq:exclamp}, $(\Psi_E^*\Psi)_{T,L}$, are given in Ref.~\cite{Kowalski:2006hc}.  We use the ``boosted Gaussian'' vector meson wave functions \cite{Nemchik:1994fp,Nemchik:1996cw,Forshaw:2003ki}, which were found to give the best description of data in Ref.~\cite{Kowalski:2006hc}.  We take $x=\xB(1+M_V^2/Q^2)$ for vector meson production.  For DVCS we take $x=\xB$ for the light quark contributions and $x=\xB(1+4m_c^2/Q^2)$ for the charm quark contribution.

\begin{figure}
  \centering
  \includegraphics[width=0.33\textwidth]{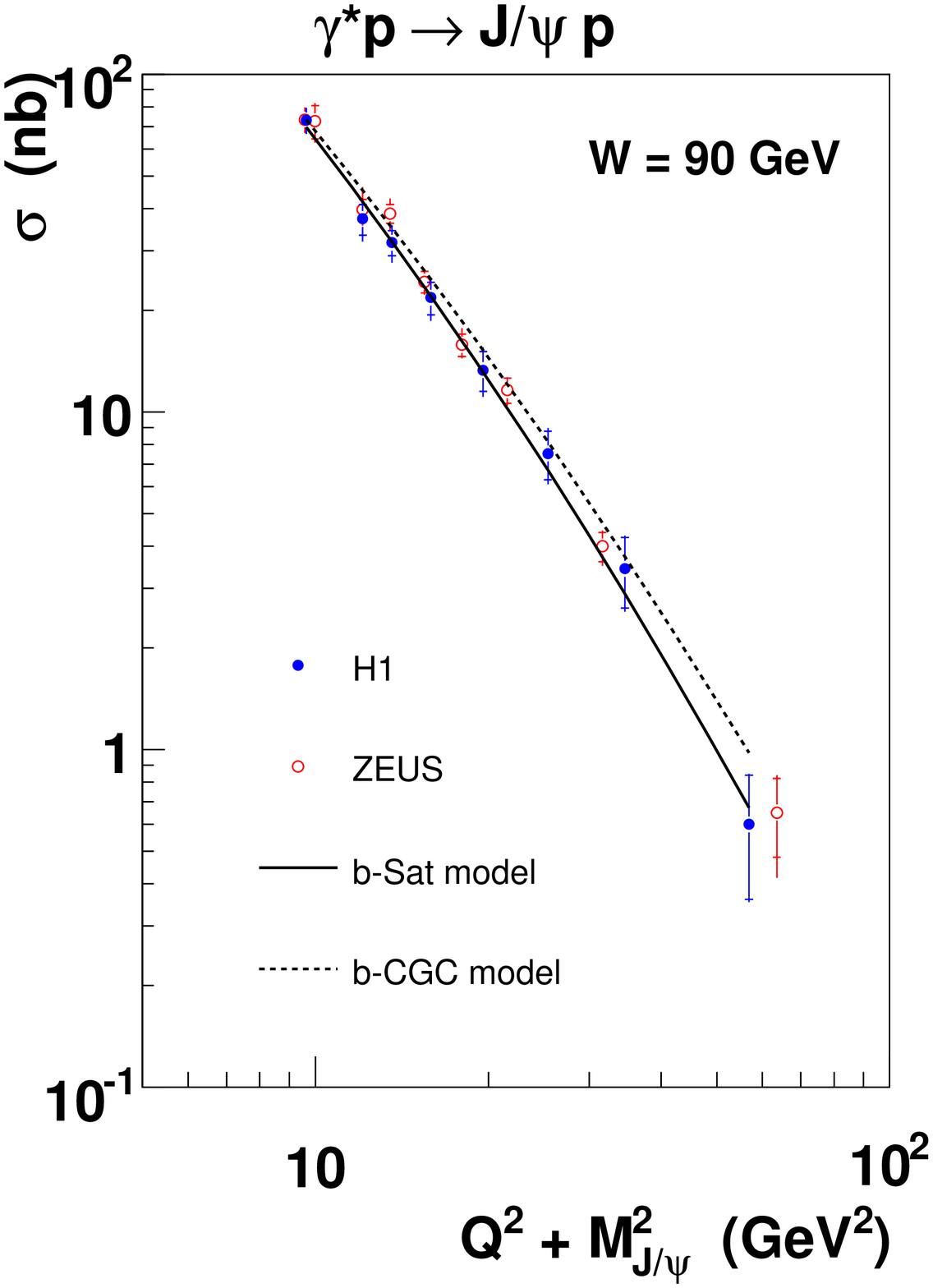}%
  \includegraphics[width=0.33\textwidth]{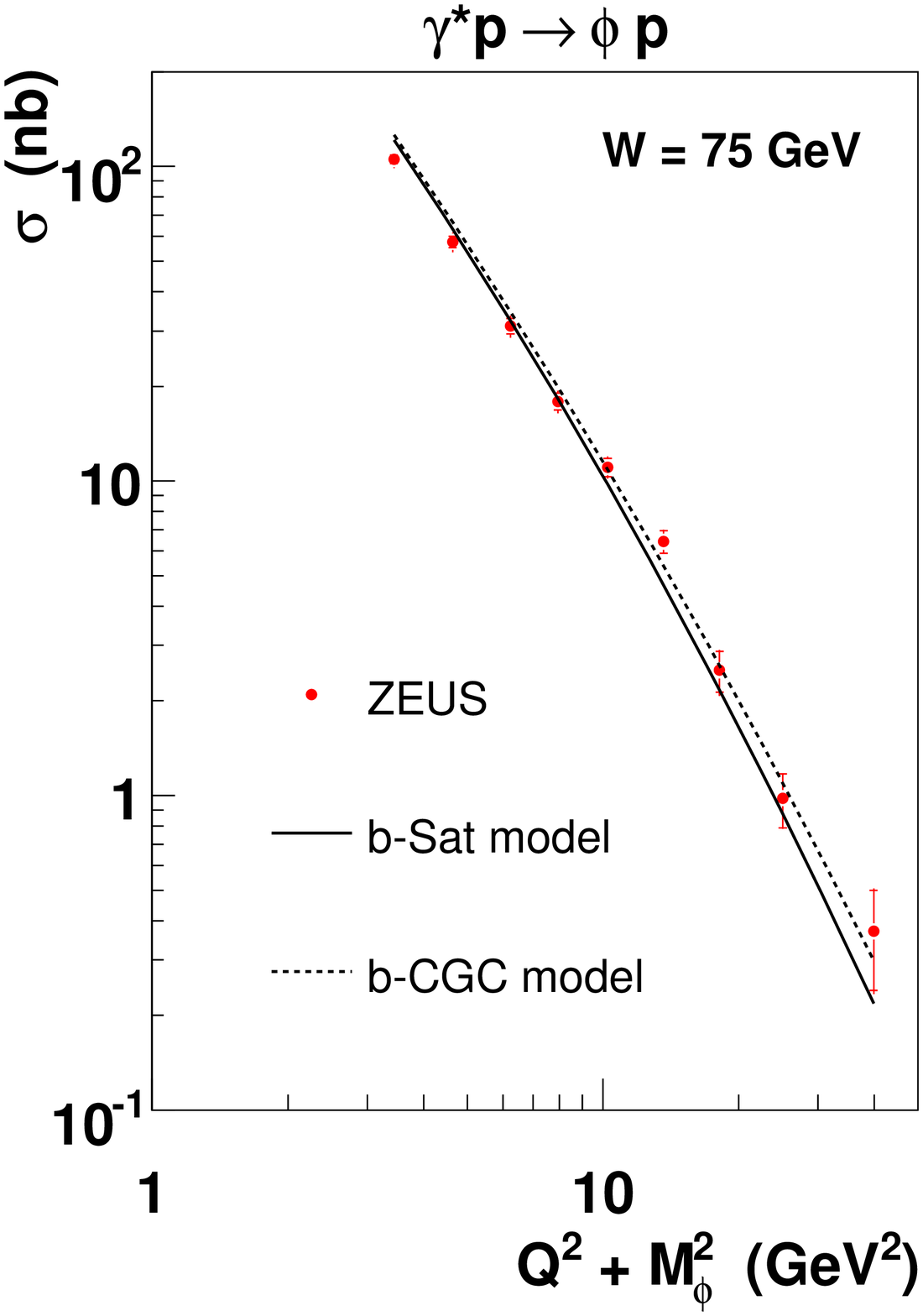}%
  \includegraphics[width=0.33\textwidth]{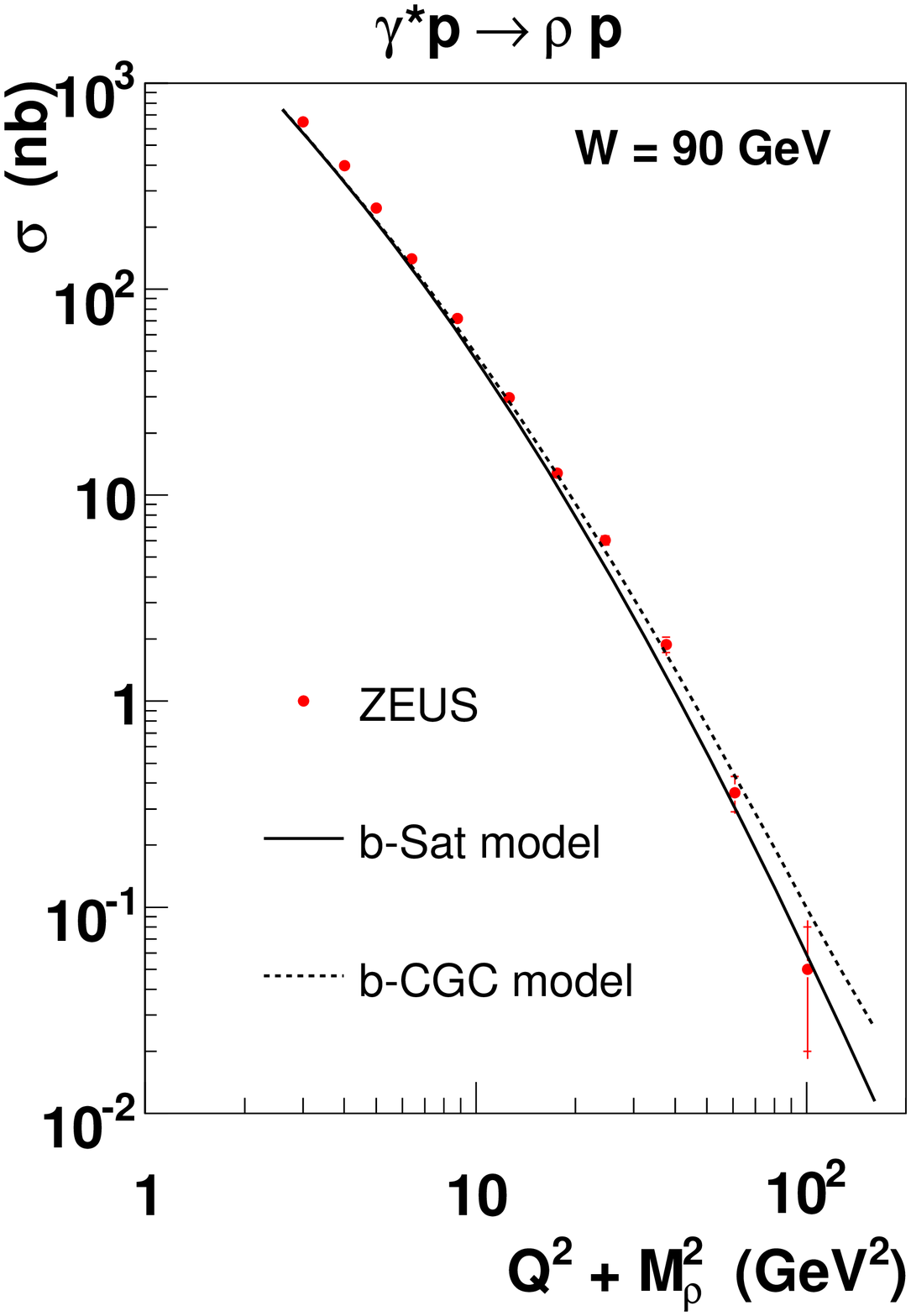}
  \caption{The total cross section $\sigma$ vs.~$(Q^2+M_V^2)$ for exclusive $J/\psi$ \cite{Chekanov:2002xi,Chekanov:2004mw,Aktas:2005xu}, $\phi$ \cite{Chekanov:2005cq} and $\rho$ \cite{Chekanov:2007zr} meson production compared to predictions from the b-Sat and b-CGC models using the ``boosted Gaussian'' vector meson wave function.  The ZEUS $J/\psi$ photoproduction point is taken from Table 1 of Ref.~\cite{Chekanov:2002xi}, from the muon decay channel with $W = 90$--$110$ GeV.}
  \label{fig:crossq}
\end{figure}

\begin{figure}
  \centering
  \includegraphics[width=0.33\textwidth]{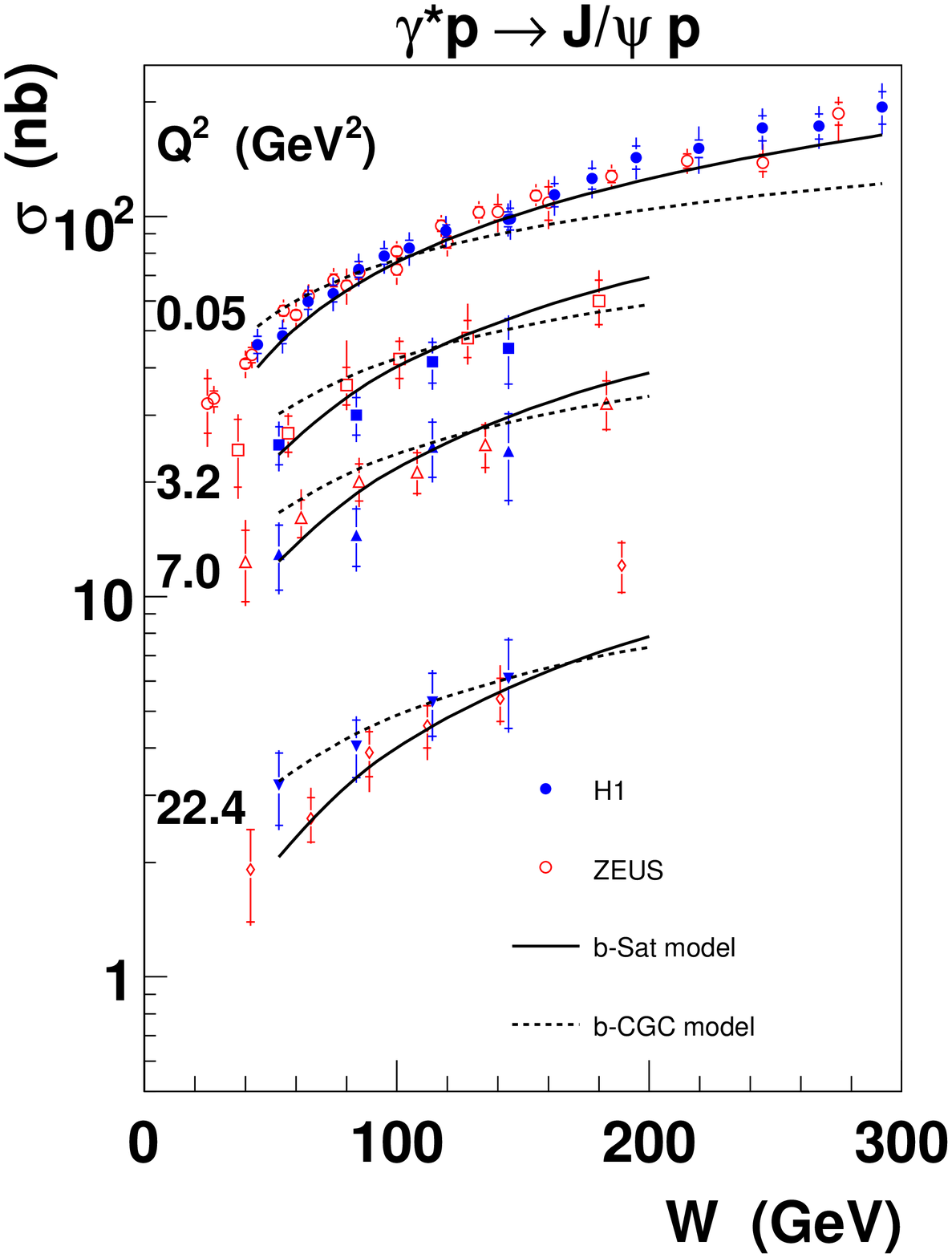}%
  \includegraphics[width=0.33\textwidth]{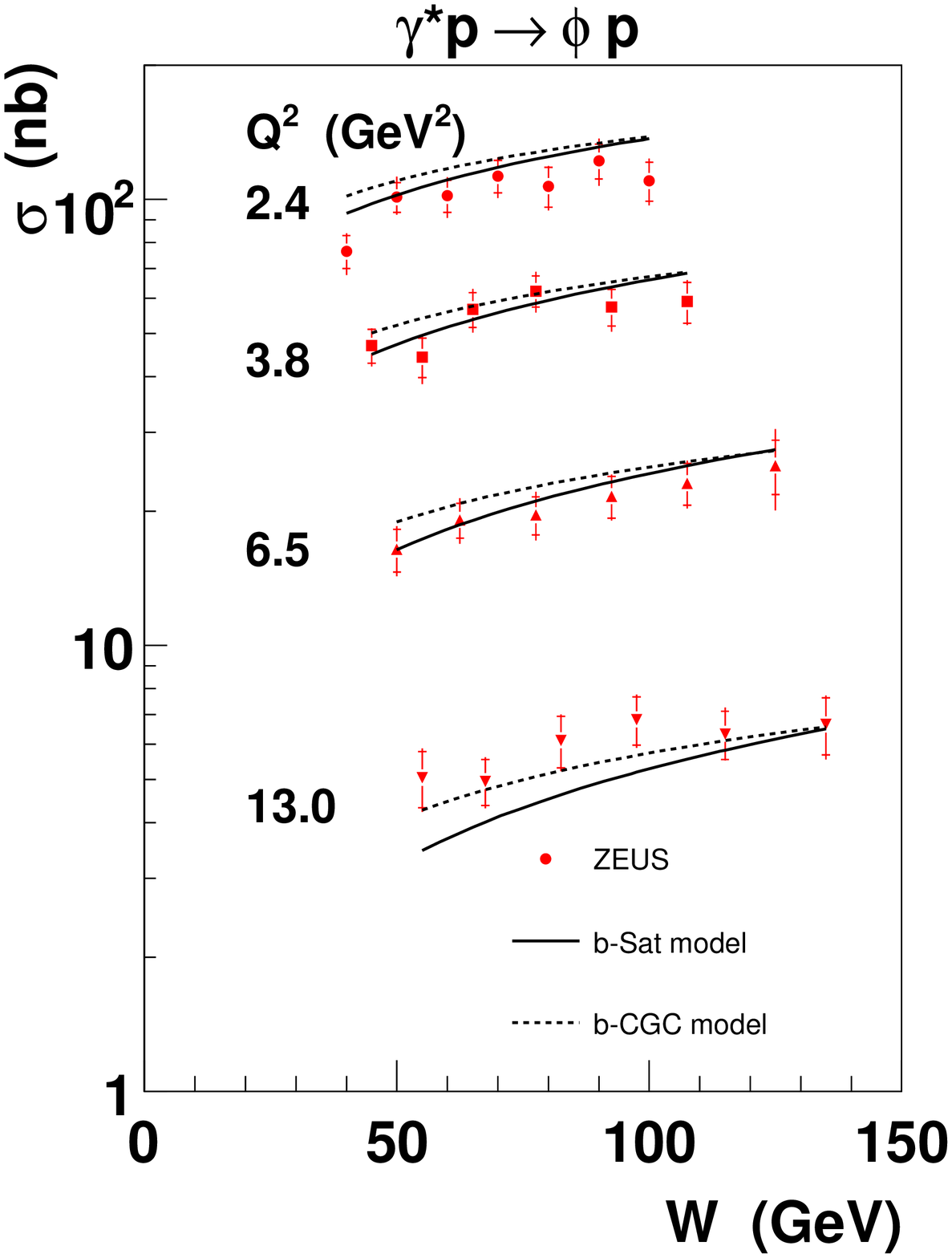}%
  \includegraphics[width=0.33\textwidth]{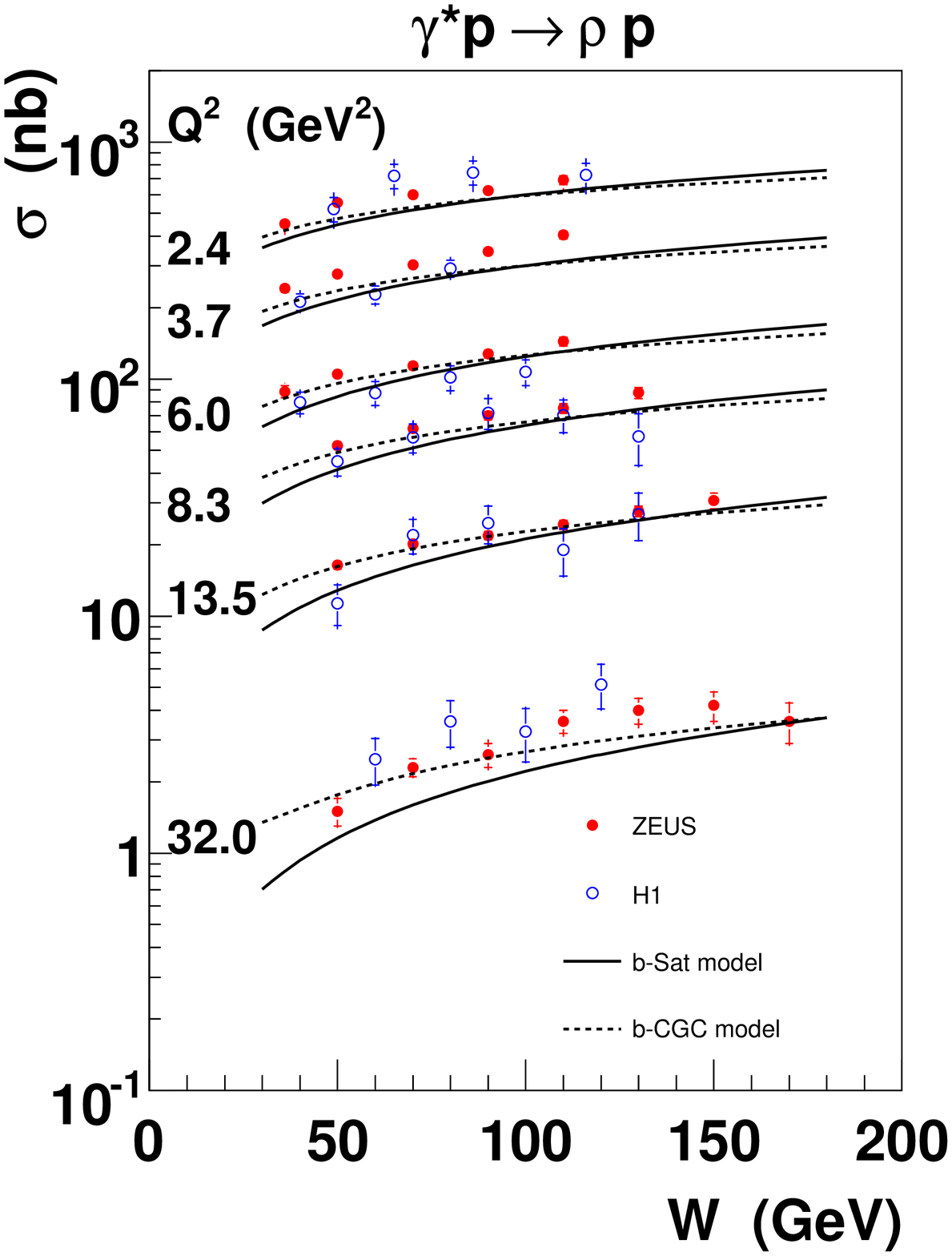}
  \caption{The total cross section $\sigma$ vs.~$W$ for exclusive $J/\psi$ \cite{Chekanov:2002xi,Chekanov:2004mw,Aktas:2005xu}, $\phi$ \cite{Chekanov:2005cq} and $\rho$ \cite{Adloff:1999kg,Chekanov:2007zr} meson production compared to predictions from the b-Sat and b-CGC models using the ``boosted Gaussian'' vector meson wave function.  The ZEUS $J/\psi$ data points \cite{Chekanov:2002xi,Chekanov:2004mw} have been scaled to the H1 $Q^2$ values \cite{Aktas:2005xu} using the $Q^2$ dependence measured by ZEUS of the form $\sigma\propto (Q^2+M_V^2)^{-2.44}$ \cite{Chekanov:2004mw}.  The H1 $\rho$ data points \cite{Adloff:1999kg} have been scaled to the ZEUS $Q^2$ values \cite{Chekanov:2007zr} using the $Q^2$ dependence measured by H1 of the form $\sigma\propto (Q^2+M_V^2)^{-2.24}$ \cite{Adloff:1999kg}.}
  \label{fig:crossw}
\end{figure}

\begin{figure}
  \centering
  \includegraphics[width=0.33\textwidth]{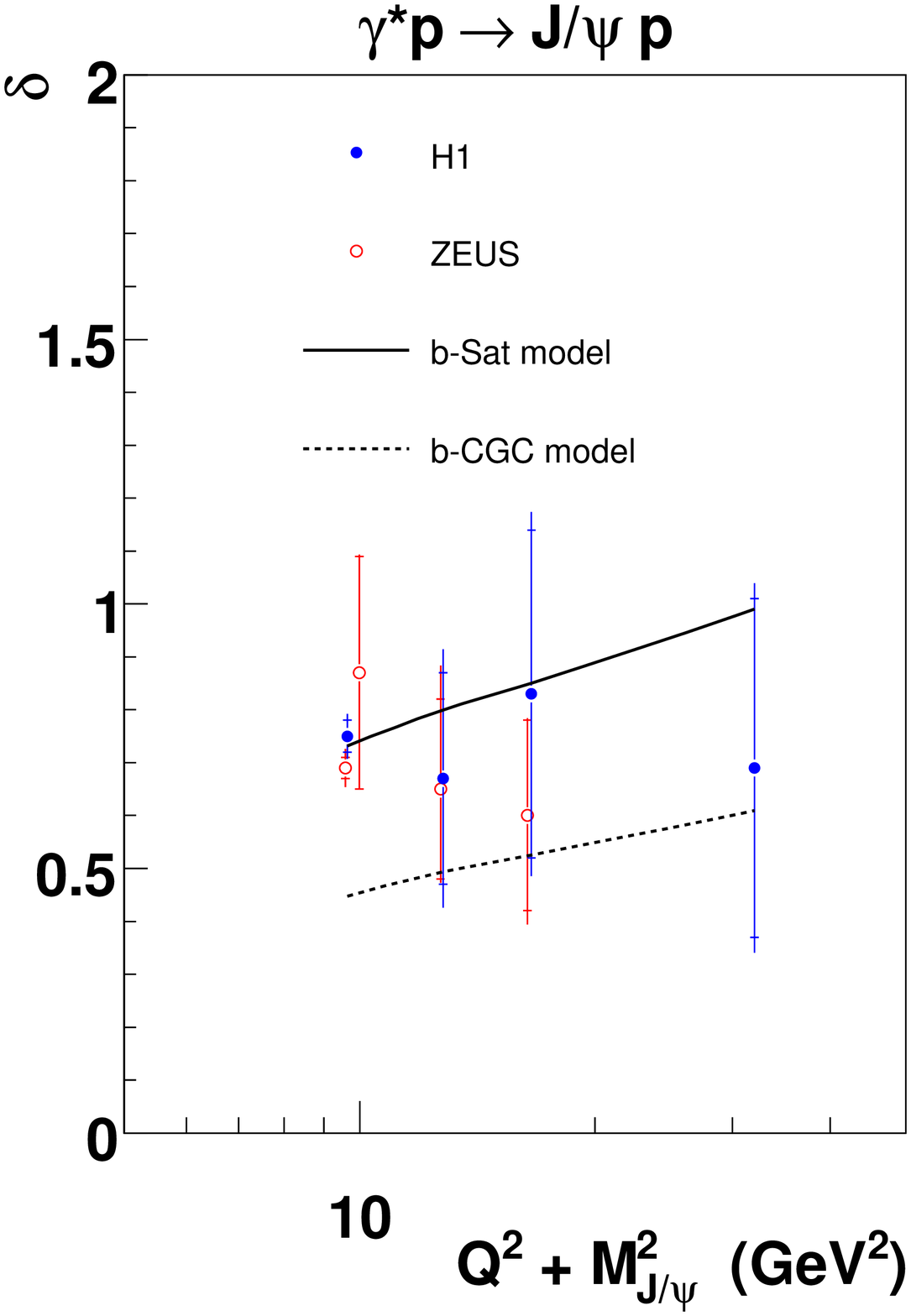}%
  \includegraphics[width=0.33\textwidth]{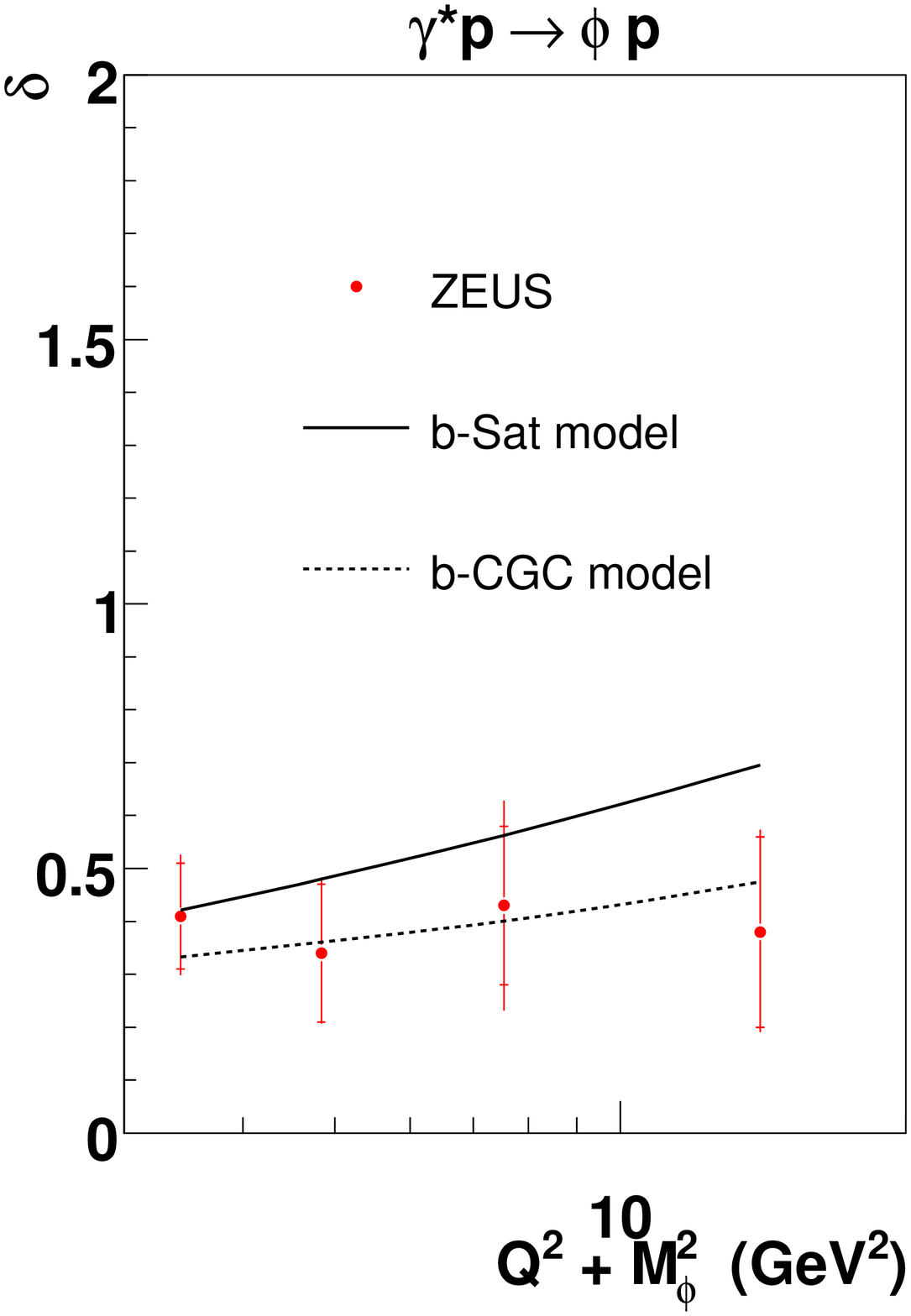}%
  \includegraphics[width=0.33\textwidth]{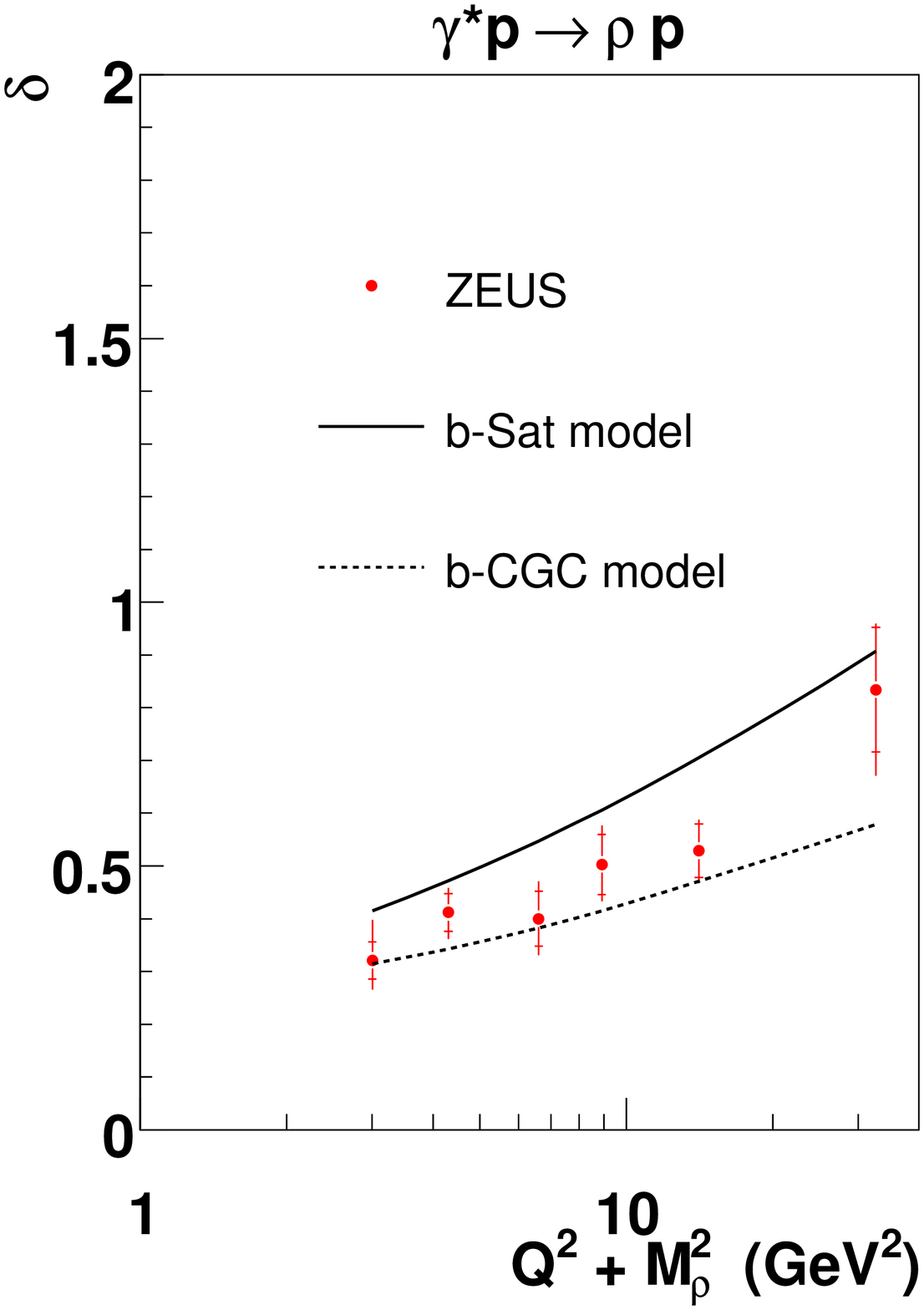}
  \caption{The power $\delta$ vs.~$(Q^2+M_V^2)$, where $\delta$ is defined by fitting $\sigma\propto W^\delta$, for exclusive $J/\psi$ \cite{Chekanov:2002xi,Chekanov:2004mw,Aktas:2005xu}, $\phi$ \cite{Chekanov:2005cq} and $\rho$ \cite{Chekanov:2007zr} meson production compared to predictions from the b-Sat and b-CGC models using the ``boosted Gaussian'' vector meson wave function.}
  \label{fig:delta}
\end{figure}

\begin{figure}
  \centering
  \includegraphics[width=0.33\textwidth]{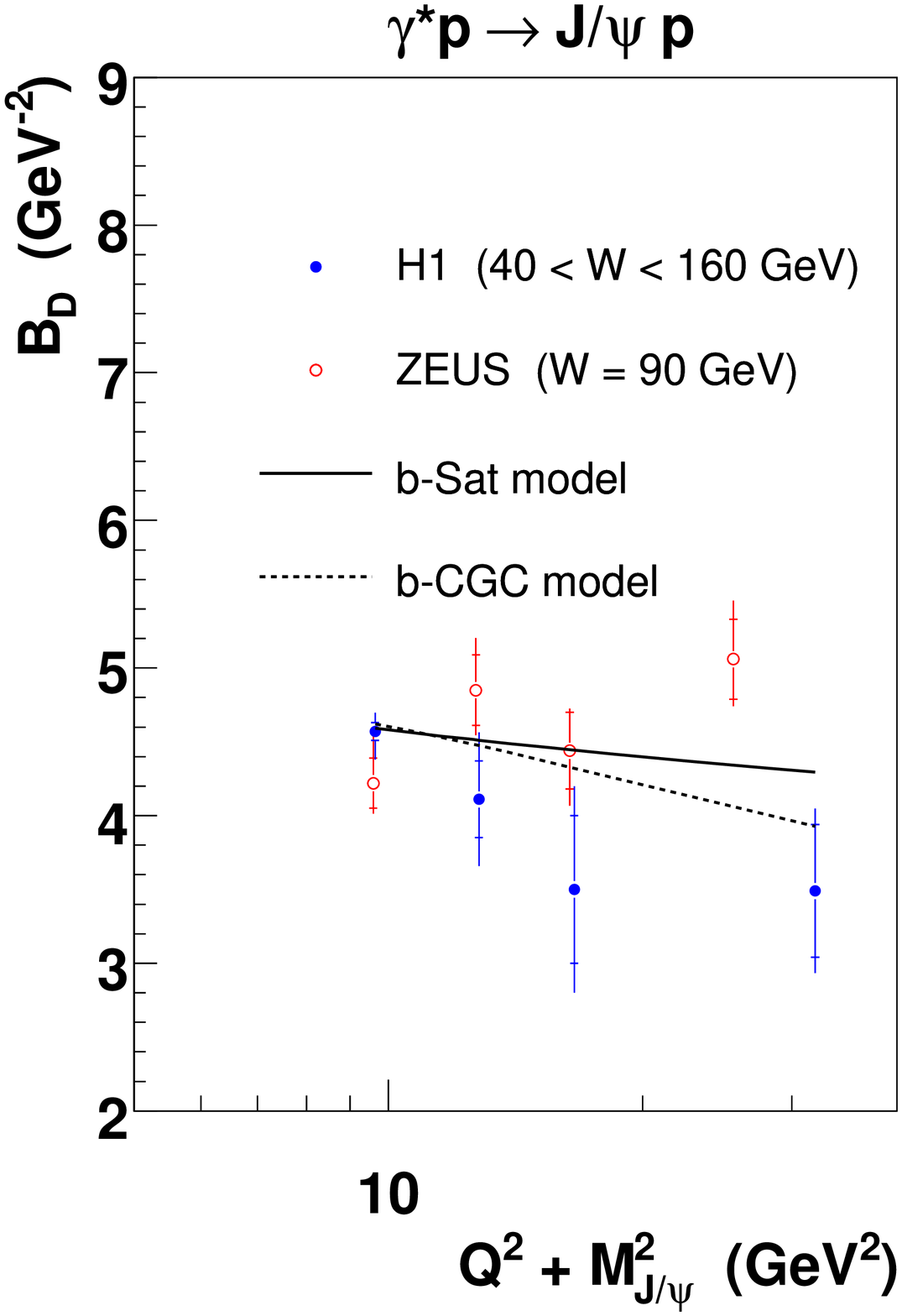}%
  \includegraphics[width=0.33\textwidth]{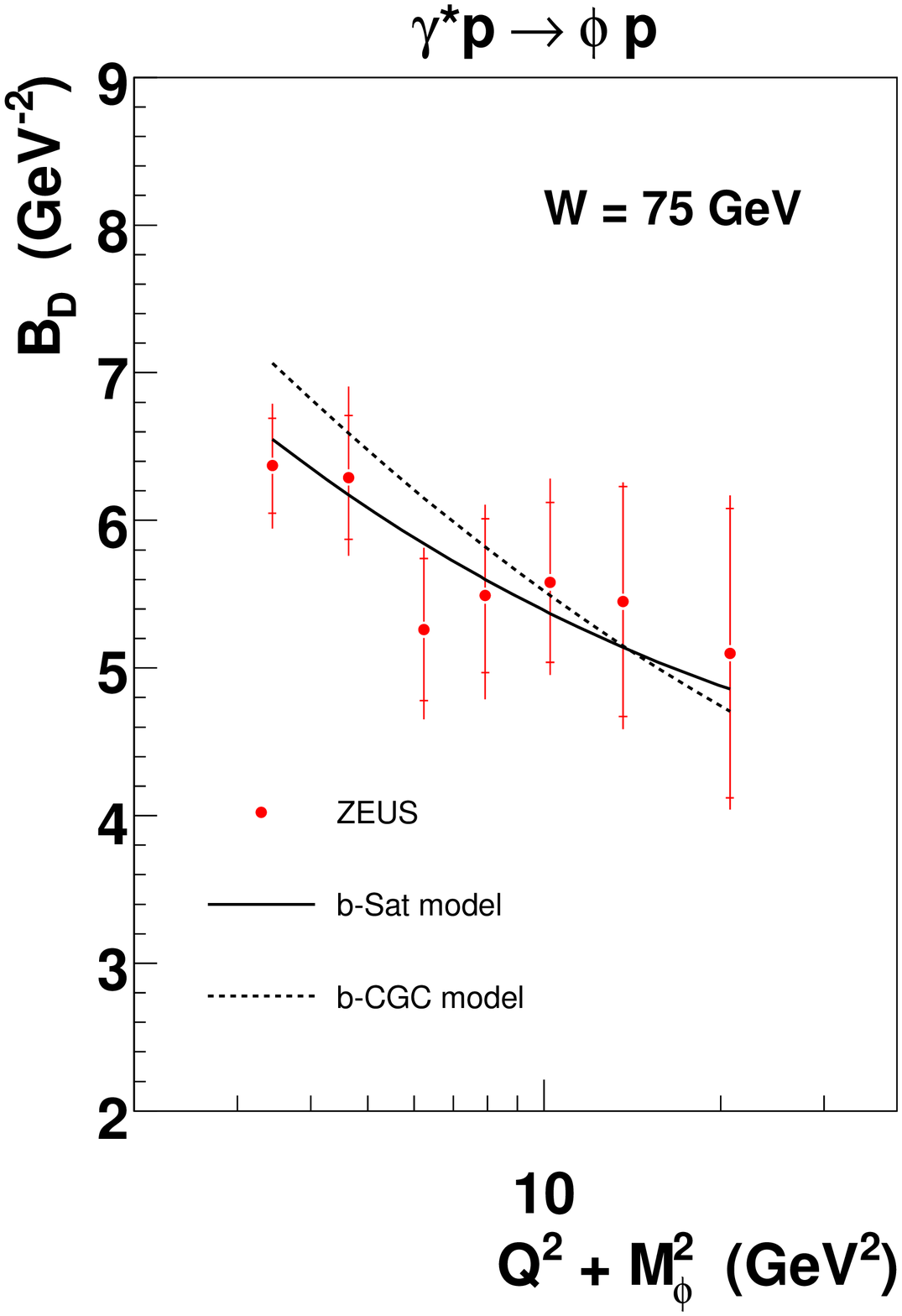}%
  \includegraphics[width=0.33\textwidth]{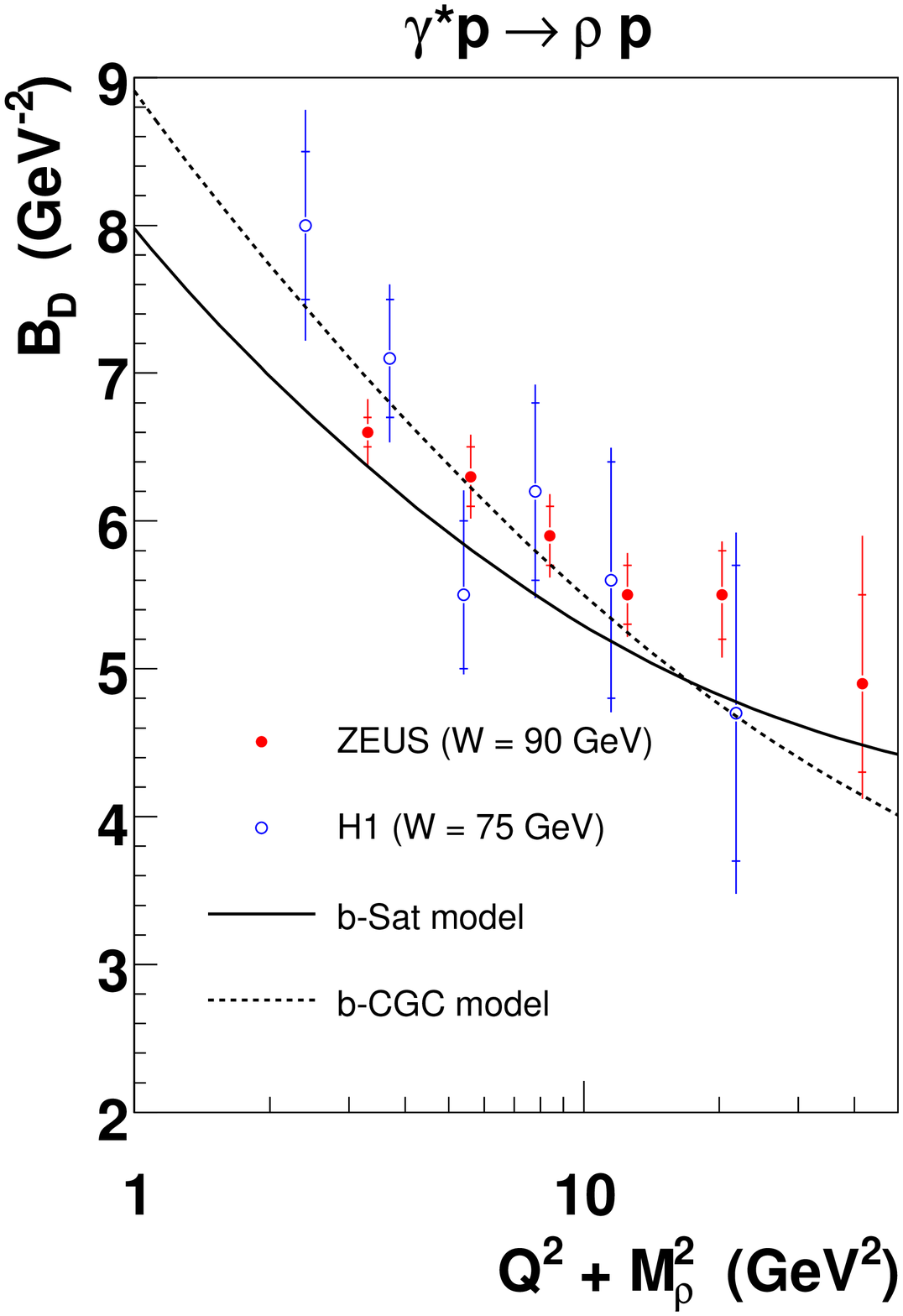}%
  \caption{The $t$-slope parameter $B_D$ vs.~$(Q^2+M_V^2)$, where $B_D$ is defined by fitting $\dif\sigma/\dif t\propto \exp(-B_D|t|)$, for exclusive $J/\psi$ \cite{Chekanov:2002xi,Chekanov:2004mw,Aktas:2005xu}, $\phi$ \cite{Chekanov:2005cq} and $\rho$ \cite{Adloff:1999kg,Chekanov:2007zr} meson production compared to predictions from the b-Sat and b-CGC models using the ``boosted Gaussian'' vector meson wave function.}
  \label{fig:bd}
\end{figure}

\begin{figure}
 \centering
  \includegraphics[width=0.33\textwidth]{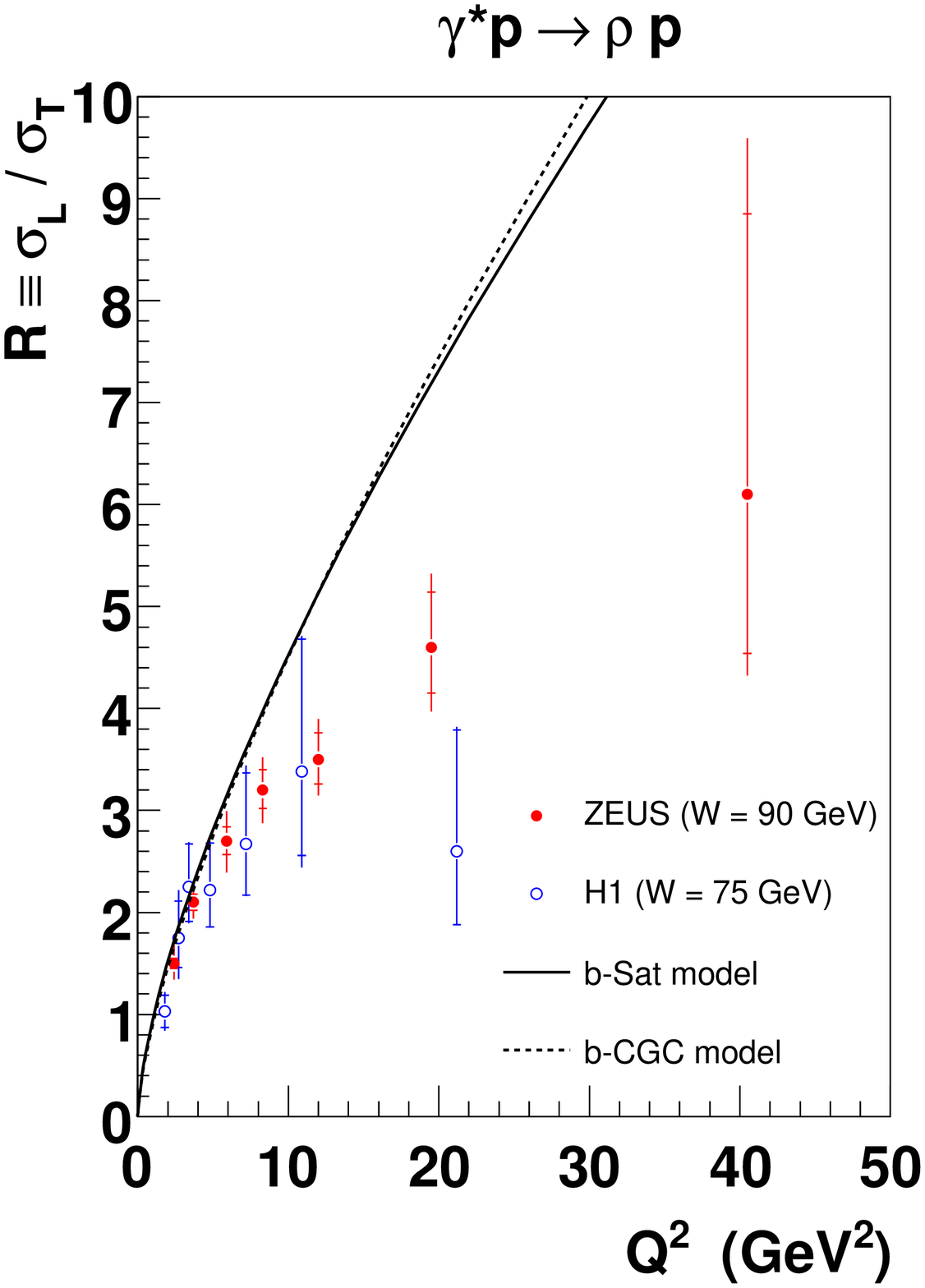}%
  \includegraphics[width=0.33\textwidth]{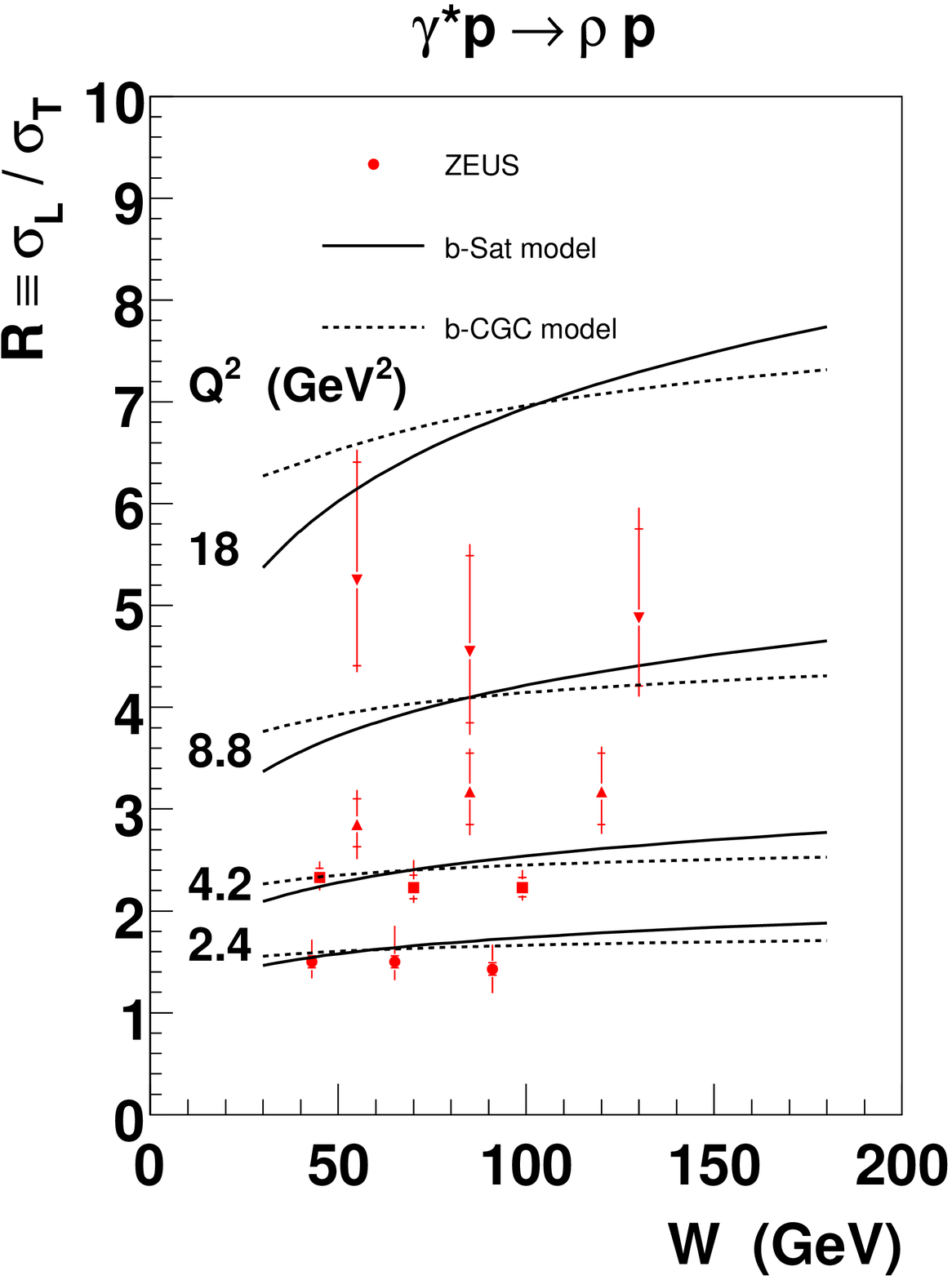}%
  \includegraphics[width=0.33\textwidth]{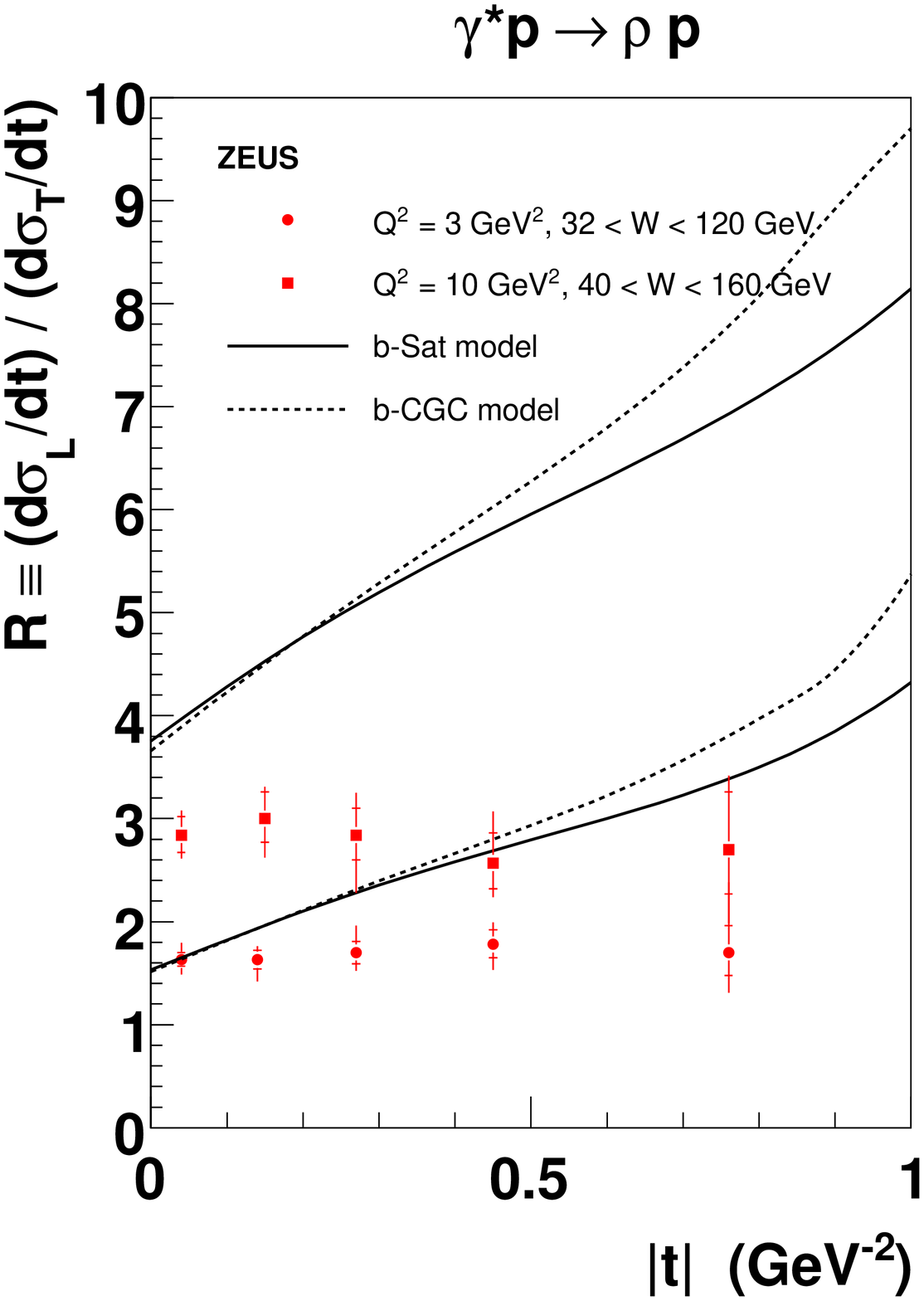}
  \caption{The ratio $R\equiv\sigma_L/\sigma_T$ vs.~$Q^2$, $W$ and $t$, for exclusive $\rho$ meson electroproduction \cite{Adloff:1999kg,Chekanov:2007zr}, compared to predictions from the b-Sat and b-CGC models using the ``boosted Gaussian'' vector meson wave function.}
  \label{fig:r}
\end{figure}

\begin{figure}
  \centering
  \includegraphics[width=0.33\textwidth]{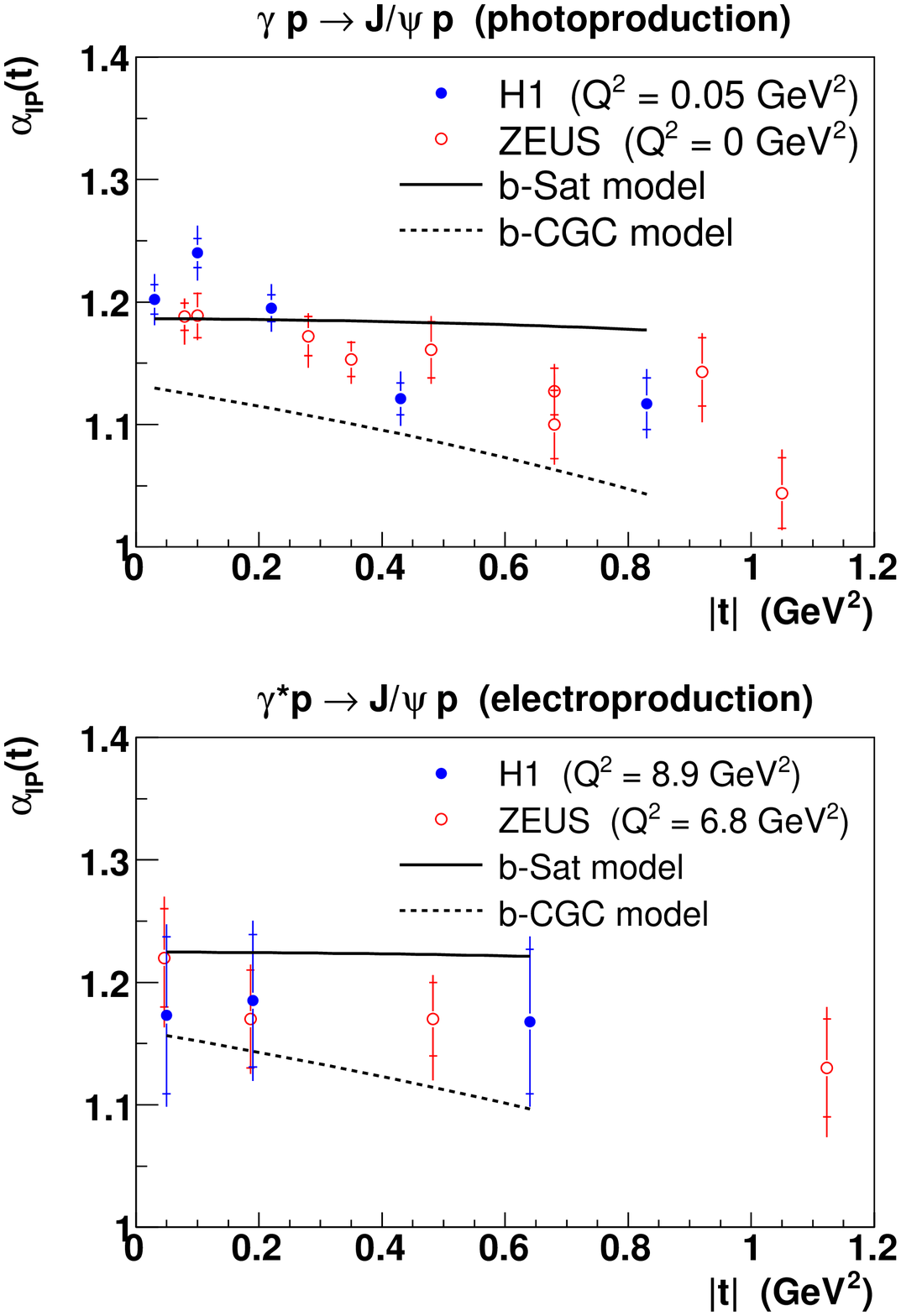}%
  \includegraphics[width=0.33\textwidth]{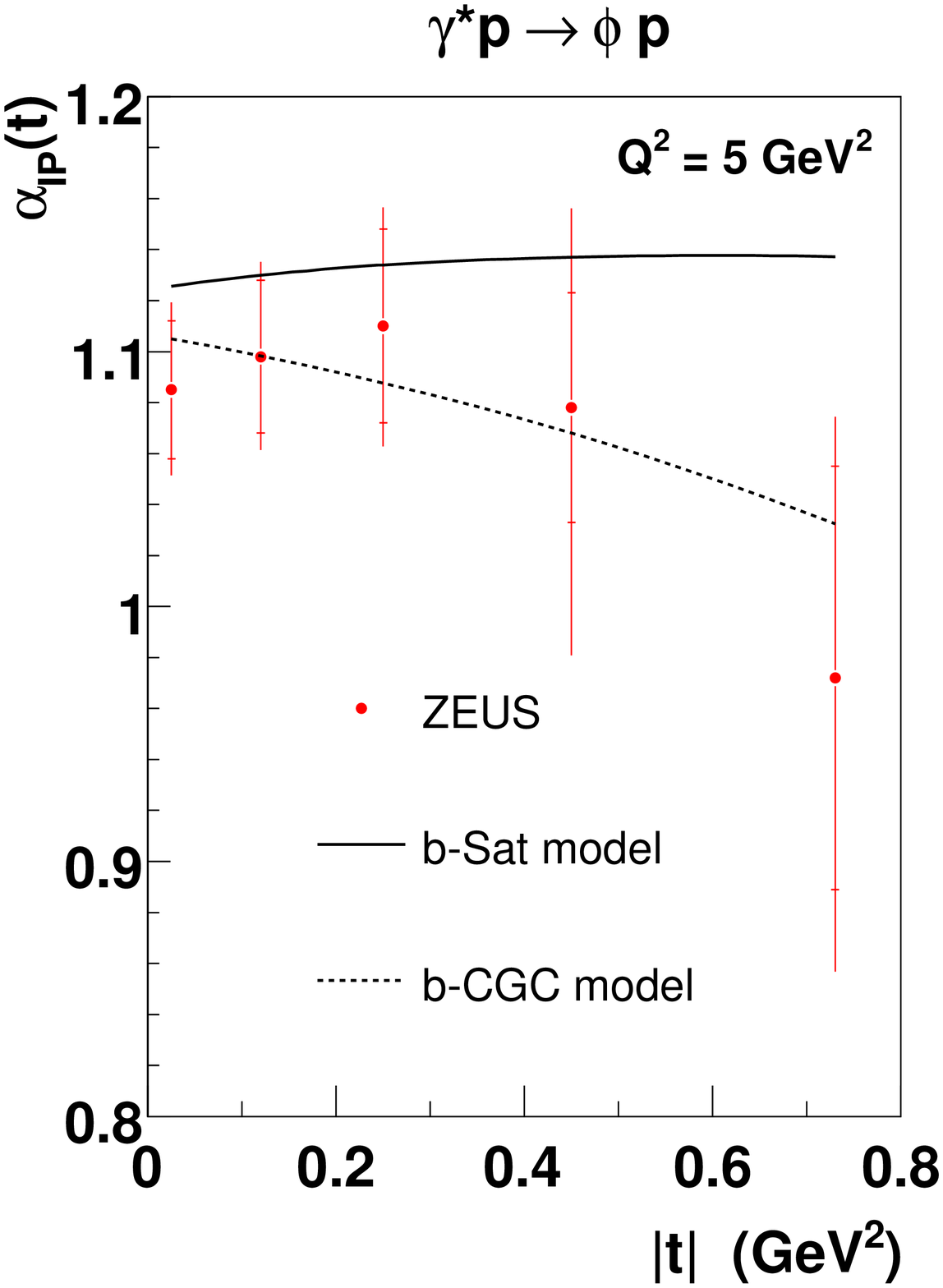}%
  \includegraphics[width=0.33\textwidth]{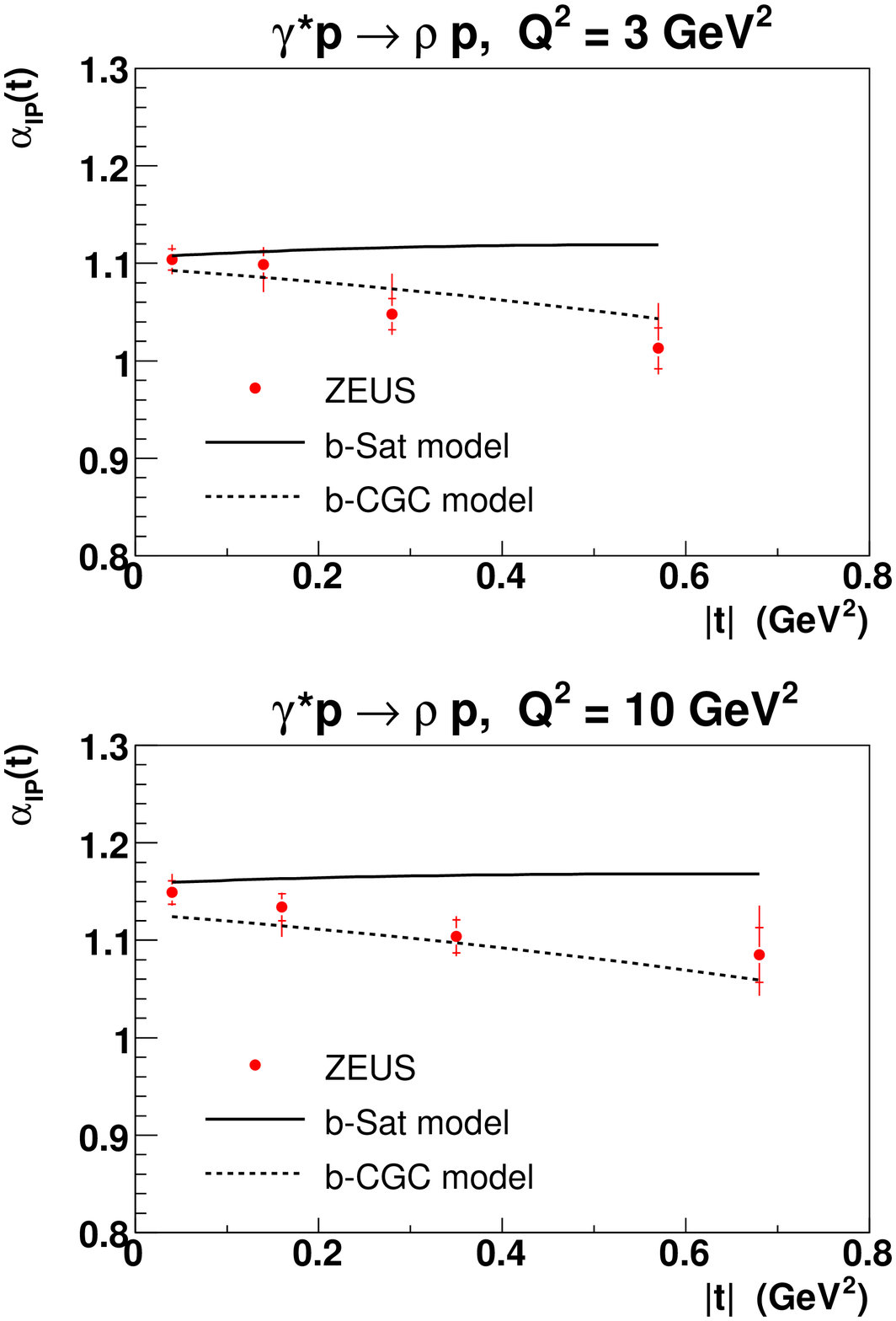}
  \caption{The effective Pomeron trajectory $\alpha_\Pom(t)$ vs.~$|t|$, where $\alpha_\Pom(t)$ is determined by fitting $\dif\sigma/\dif t\propto W^{4[\alpha_\Pom(t)-1]}$, for exclusive $J/\psi$ \cite{Chekanov:2002xi,Chekanov:2004mw,Aktas:2005xu}, $\phi$ \cite{Chekanov:2005cq} and $\rho$ \cite{Chekanov:2007zr} meson production compared to predictions from the b-Sat and b-CGC models using the ``boosted Gaussian'' vector meson wave function.}
  \label{fig:apom}
\end{figure}

\begin{figure}
  \centering
  \includegraphics[width=0.5\textwidth]{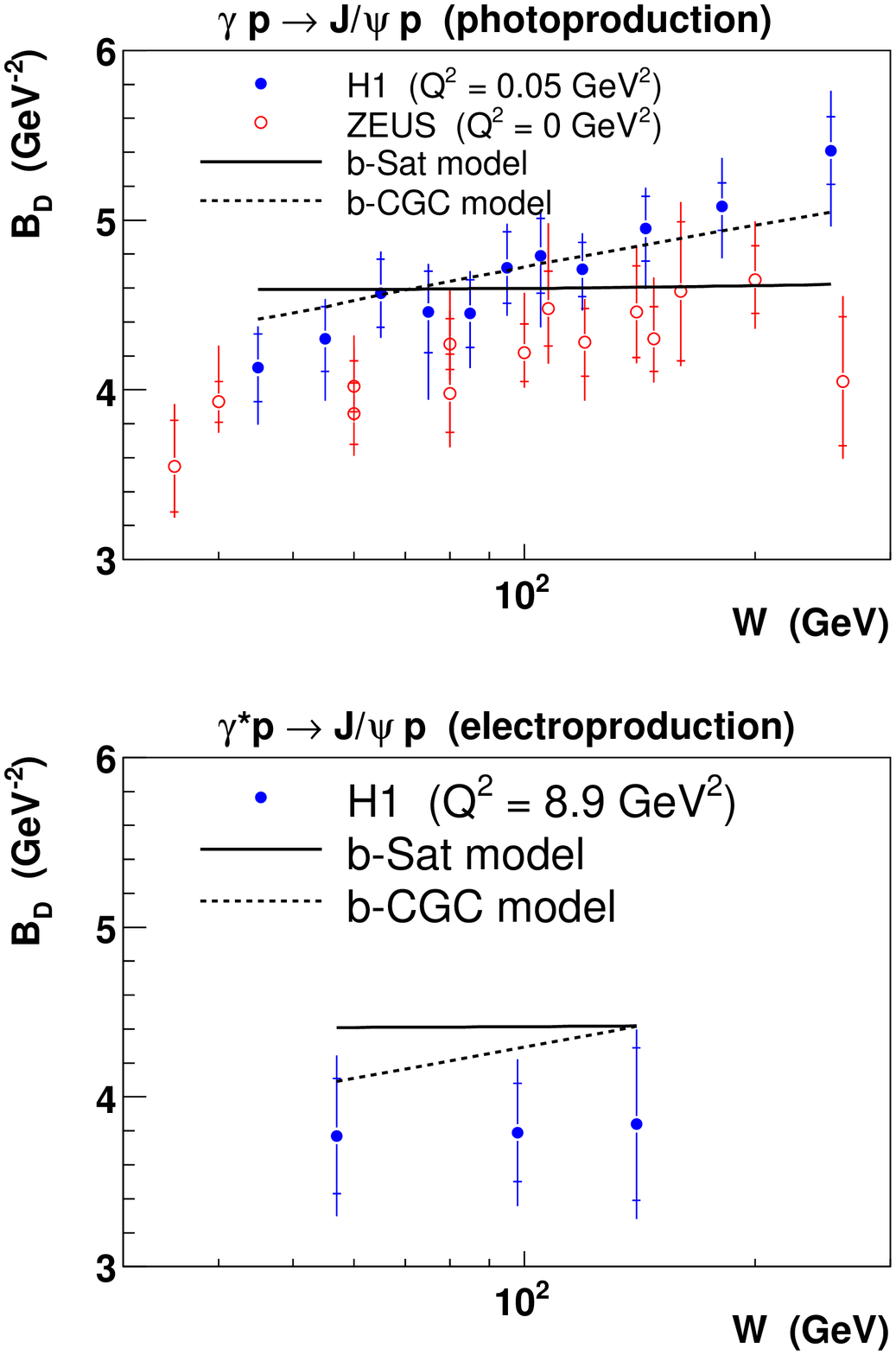}%
  \includegraphics[width=0.5\textwidth]{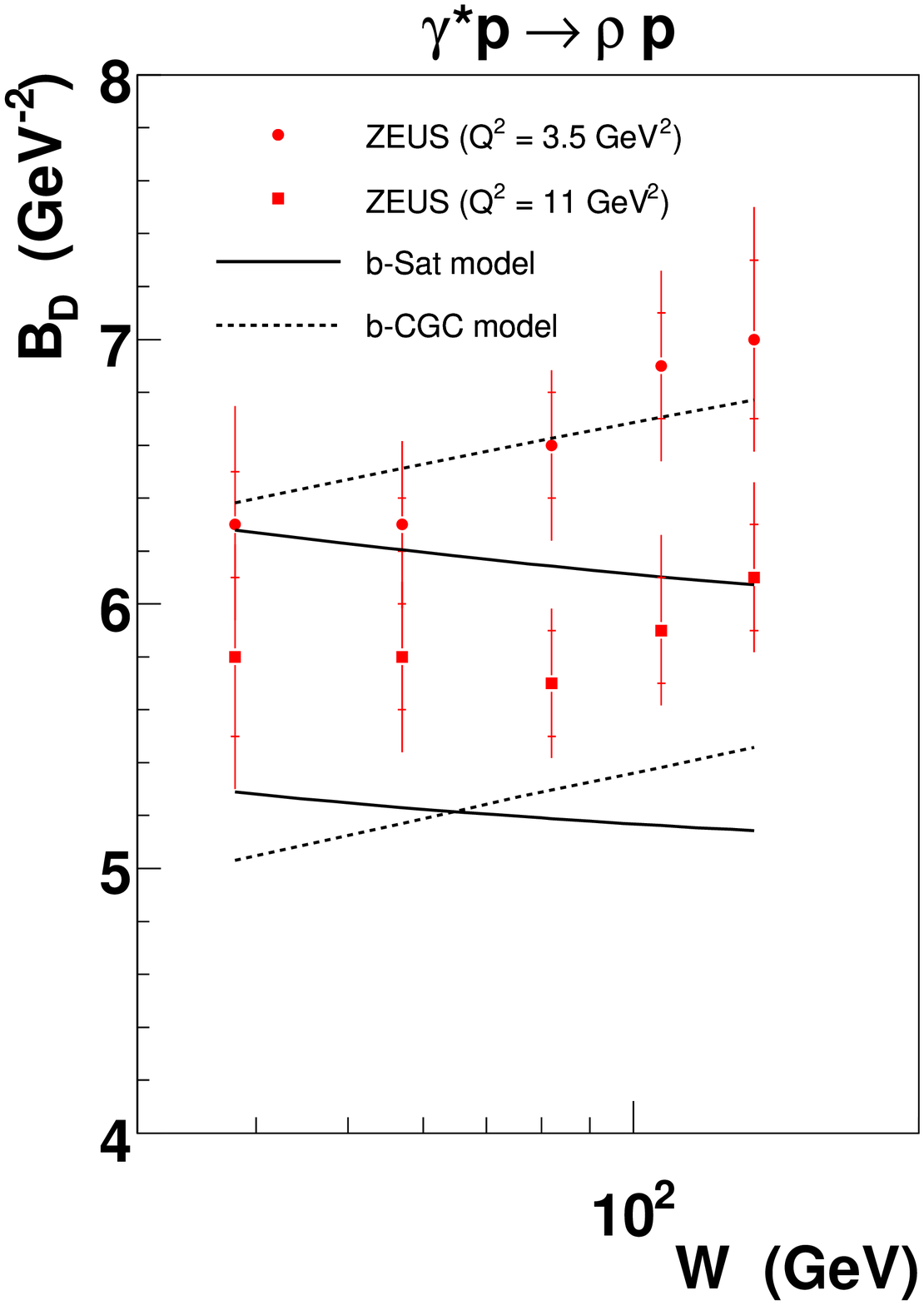}
  \caption{The $t$-slope parameter $B_D$ vs.~$W$, where $B_D$ is defined by fitting $\dif\sigma/\dif t\propto \exp(-B_D|t|)$, for exclusive $J/\psi$ \cite{Chekanov:2002xi,Aktas:2005xu} and $\rho$ \cite{Chekanov:2007zr} meson production compared to predictions from the b-Sat and b-CGC models using the ``boosted Gaussian'' vector meson wave function.}
  \label{fig:bdw}
\end{figure}

\begin{figure}
  \centering
  \includegraphics[width=0.33\textwidth]{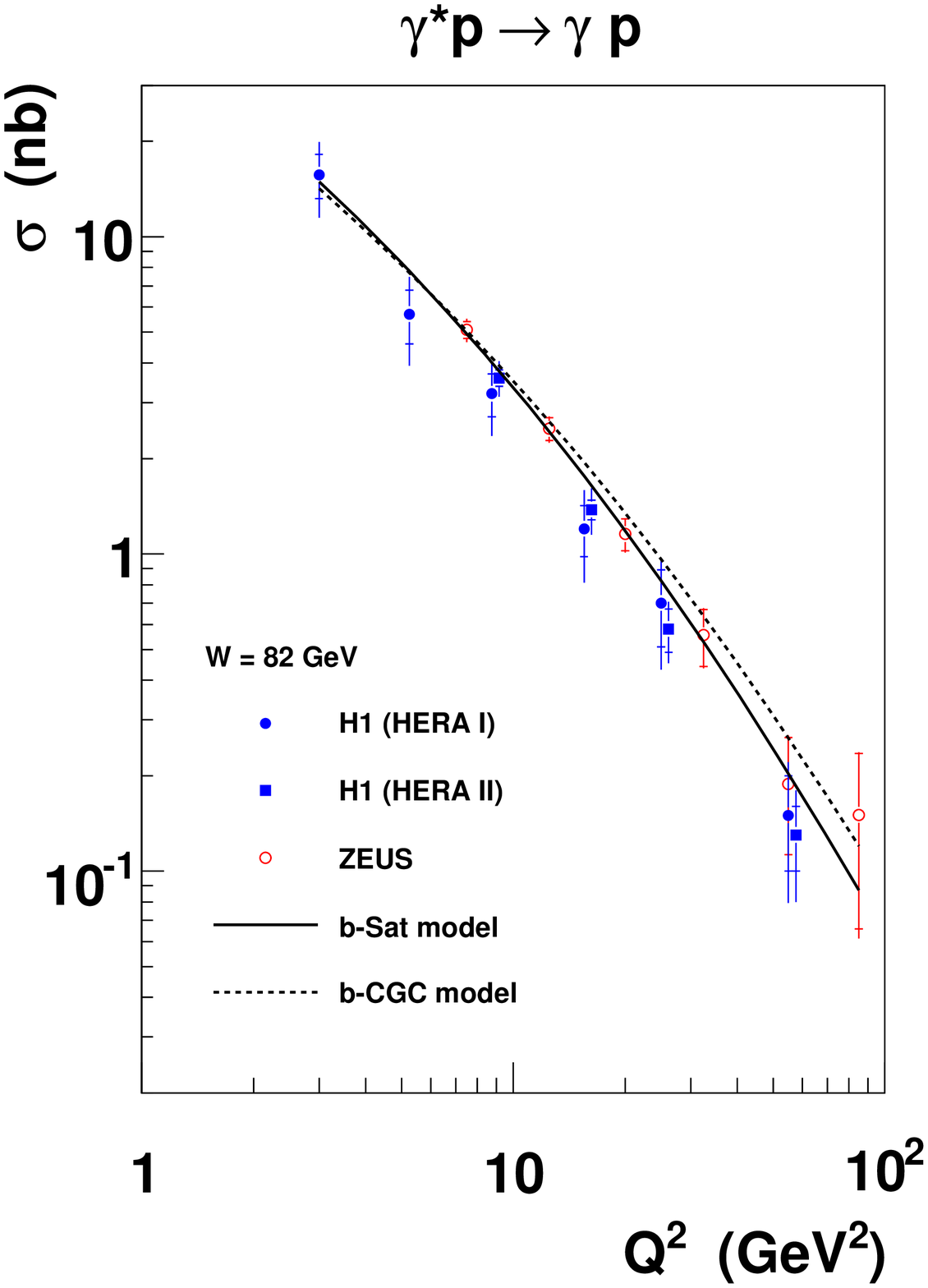}%
  \includegraphics[width=0.33\textwidth]{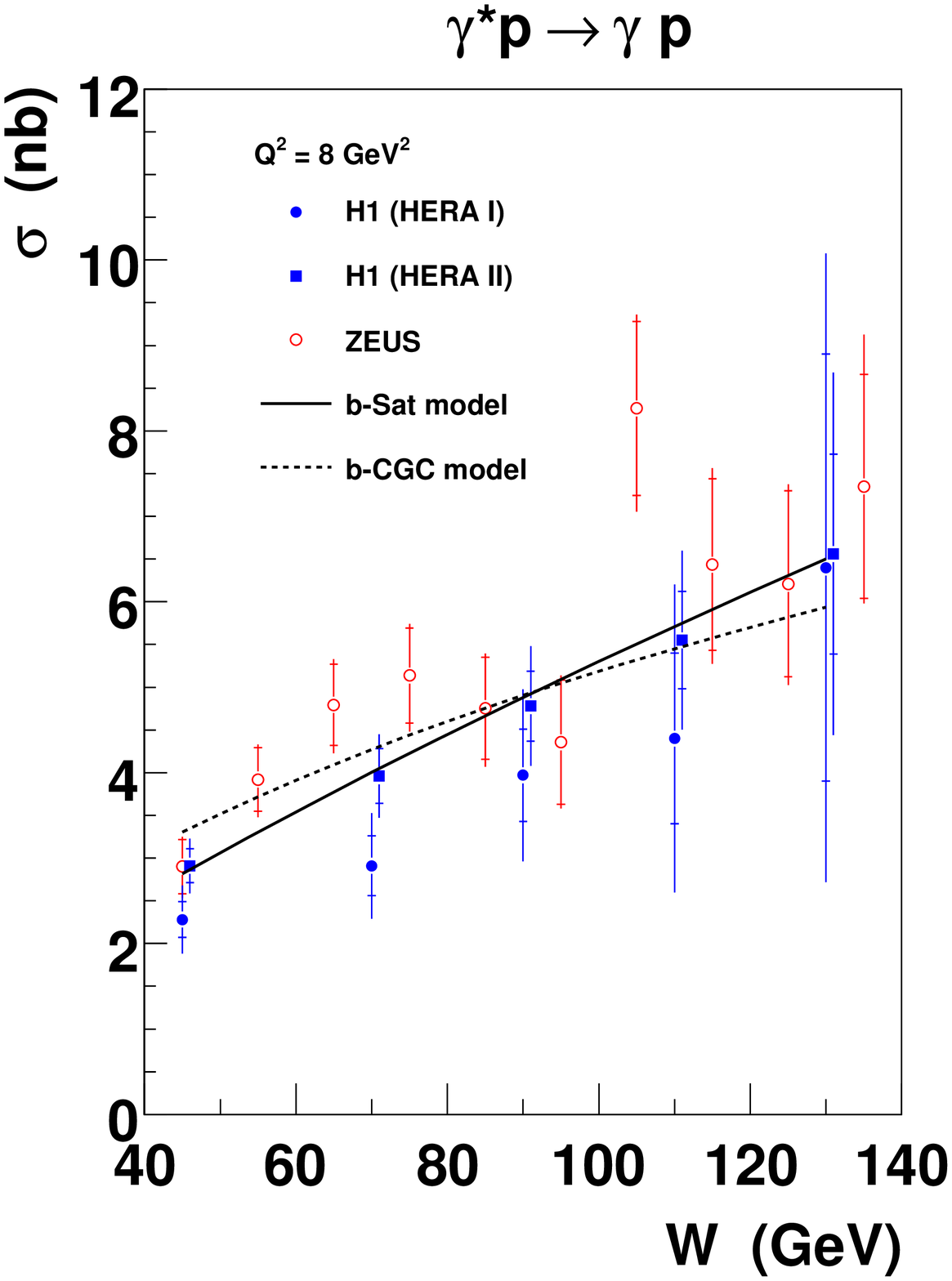}%
  \includegraphics[width=0.33\textwidth]{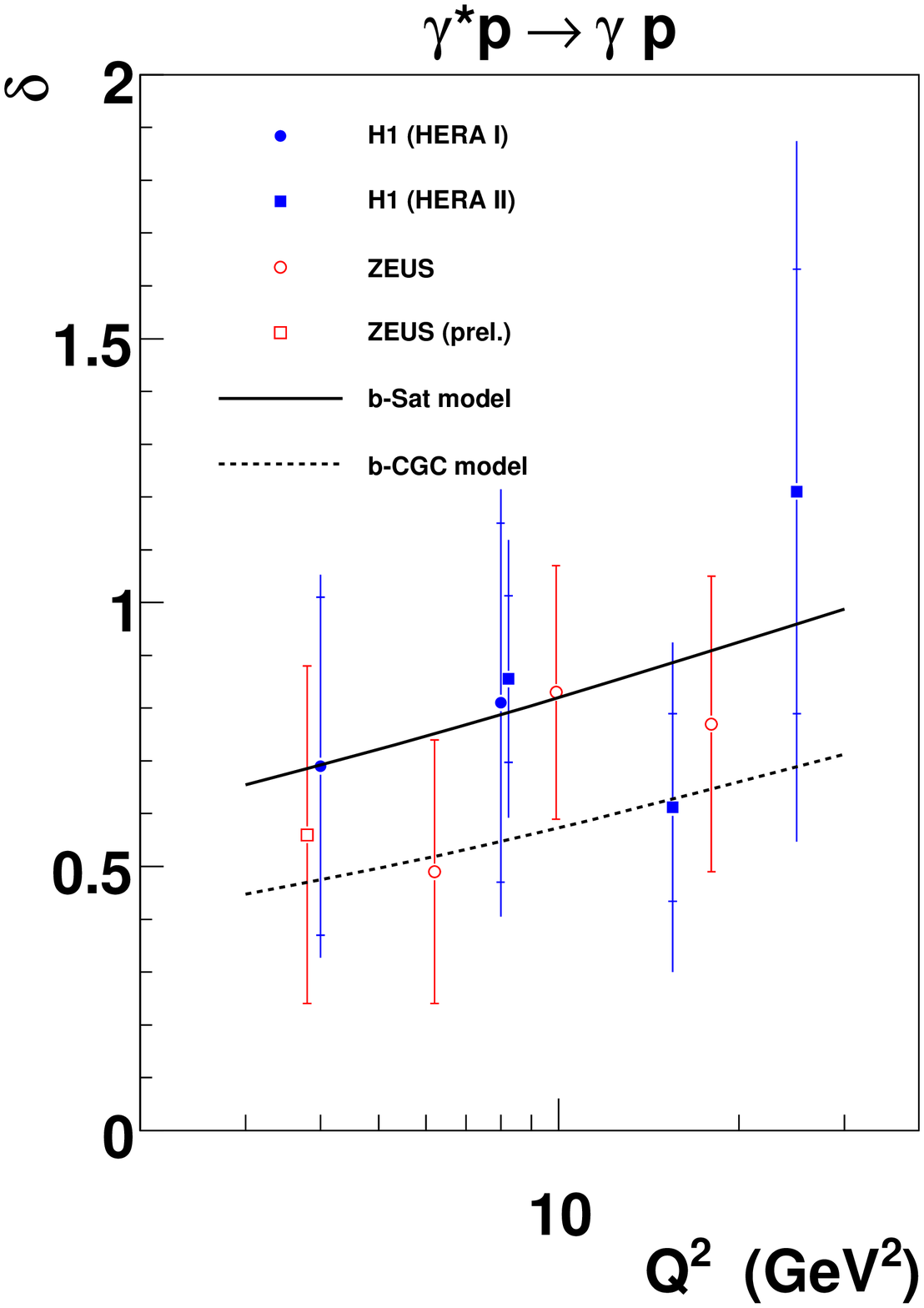}
  \caption{The total cross section $\sigma$ vs.~$Q^2$ and $W$, and the power $\delta$ vs.~$Q^2$, where $\delta$ is defined by fitting $\sigma\propto W^\delta$, for DVCS data \cite{Aktas:2005ty,Chekanov:2003ya,Aaron:2007cz,ZEUSDVCS}, compared to predictions from the b-Sat and b-CGC models.}
  \label{fig:dvcstotal}
\end{figure}

\begin{figure}
  \centering
  \hspace{0.05\textwidth}(a)\hspace{0.4\textwidth}(b)\hspace*{0.4\textwidth}\ \\
  \includegraphics[width=0.43\textwidth]{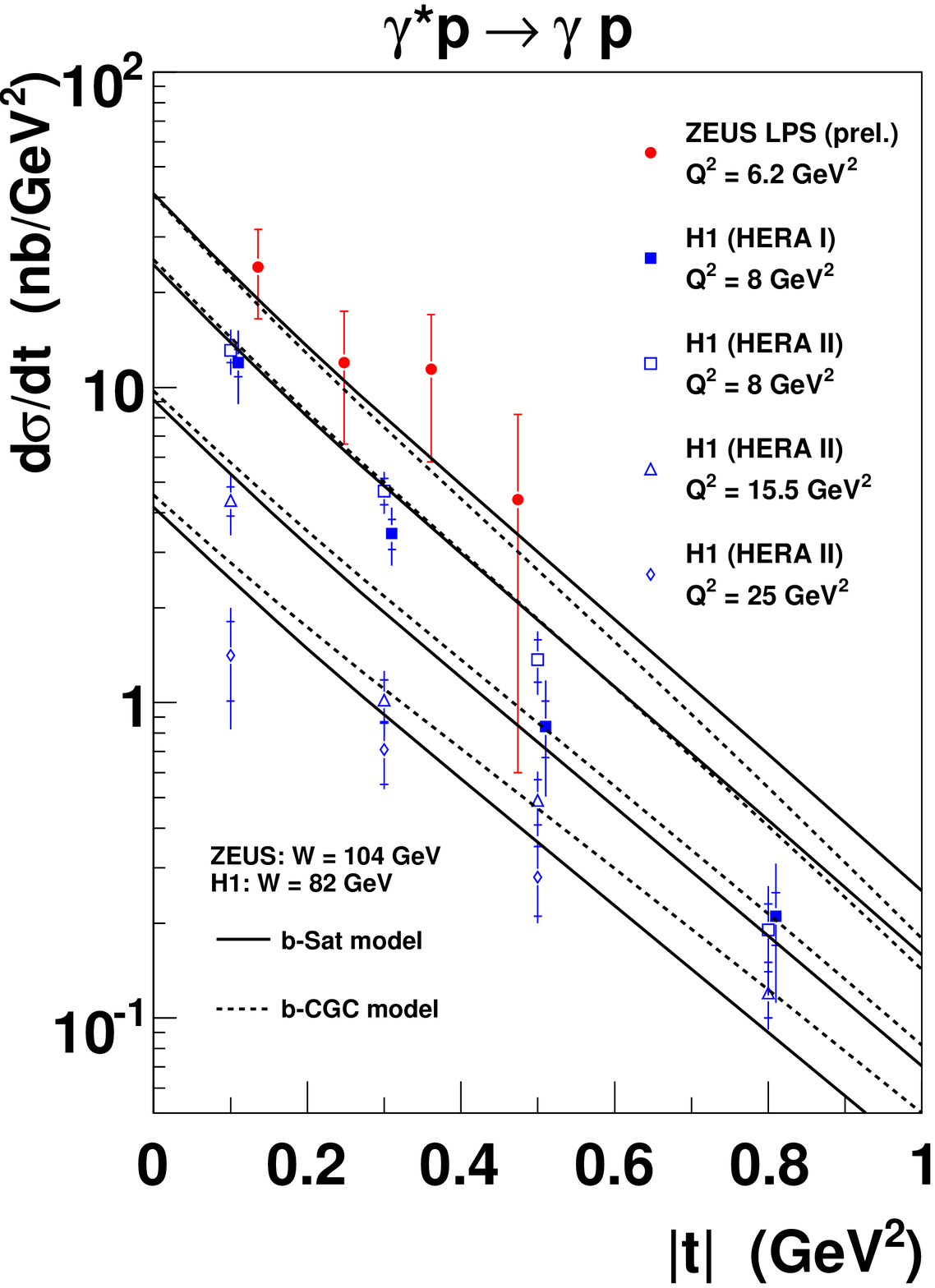}%
  \includegraphics[width=0.43\textwidth]{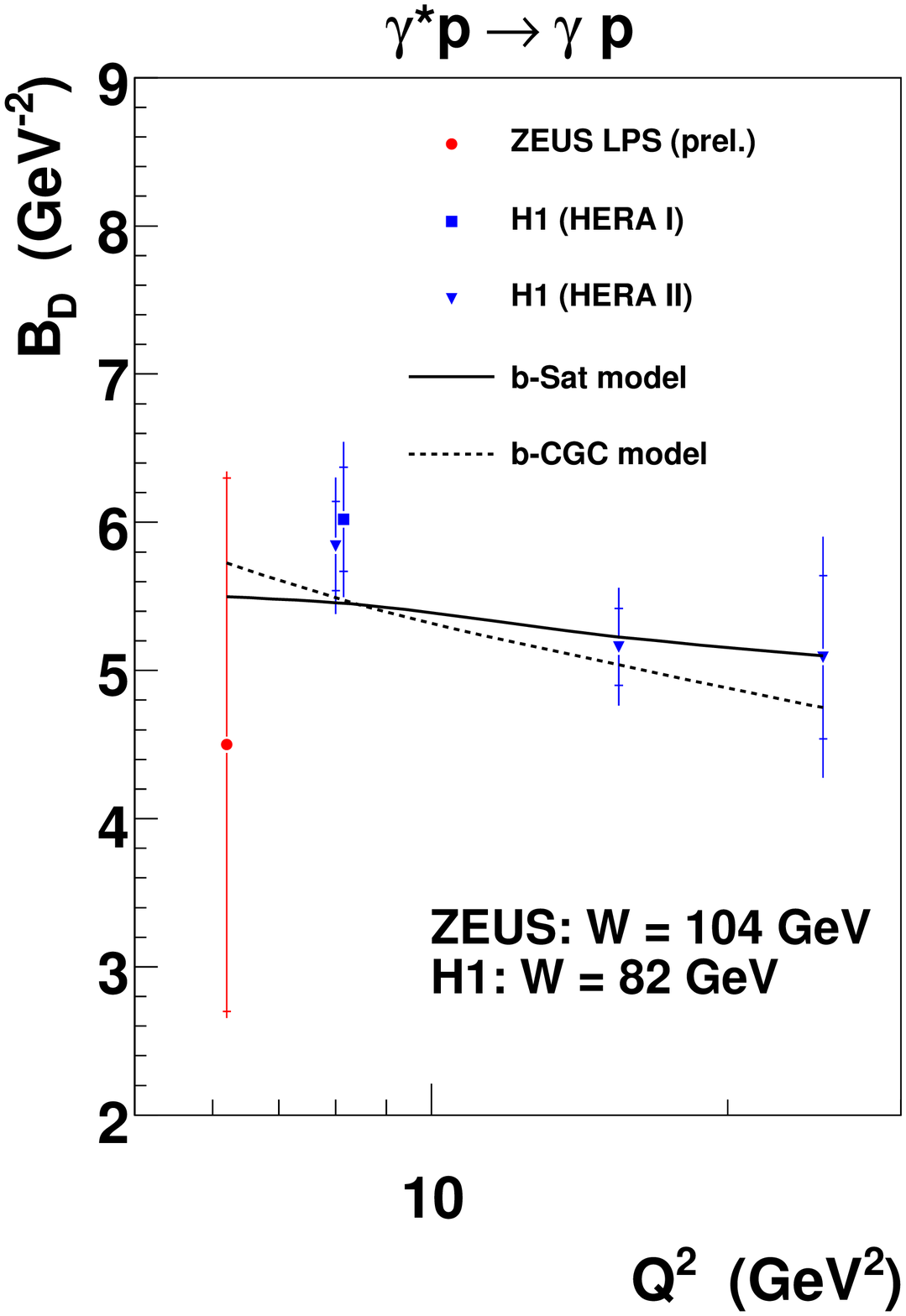}\\
  \hspace{0.05\textwidth}(c)\hspace{0.4\textwidth}(d)\hspace*{0.4\textwidth}\ \\
  \includegraphics[width=0.43\textwidth]{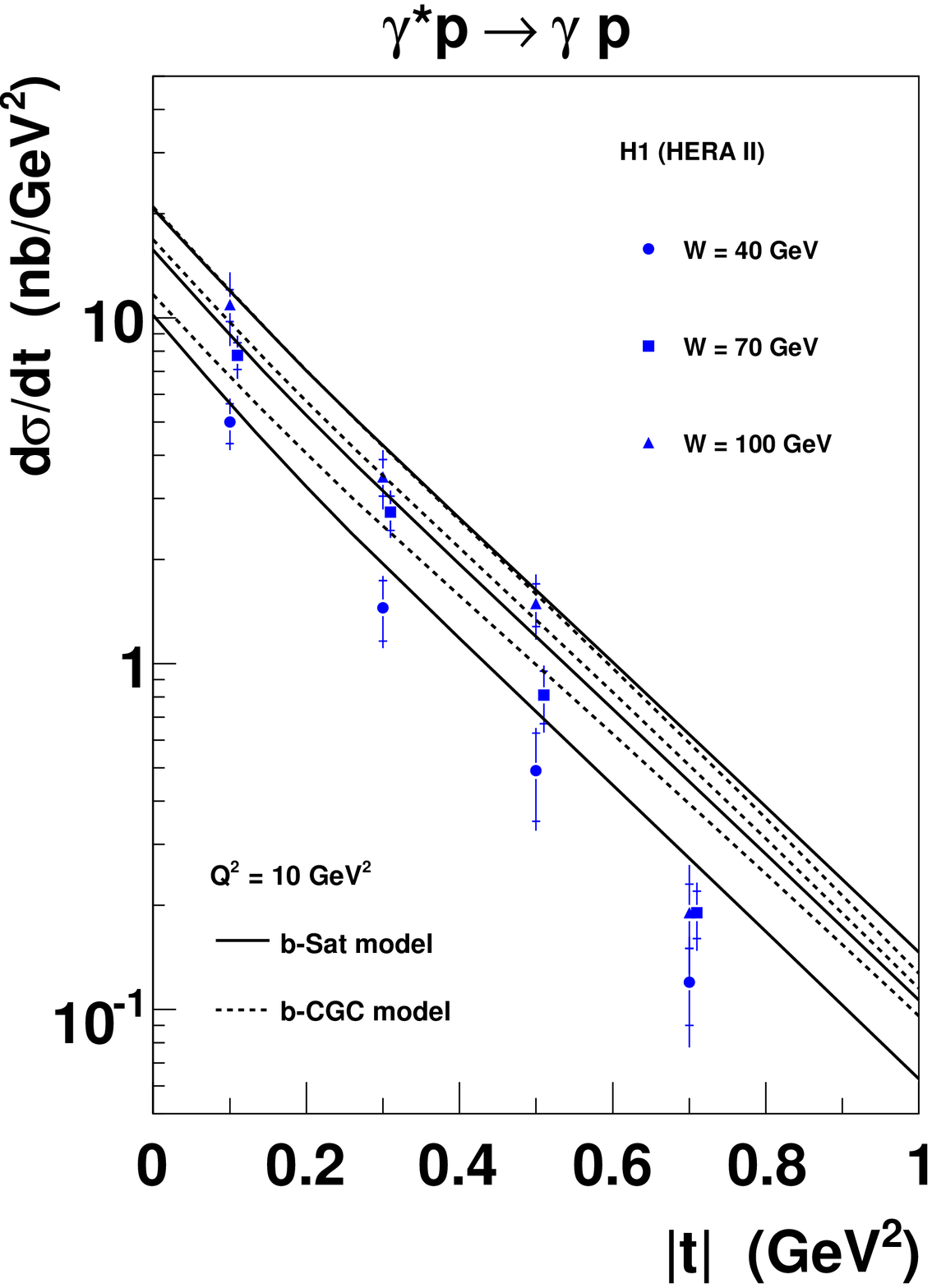}%
  \includegraphics[width=0.43\textwidth]{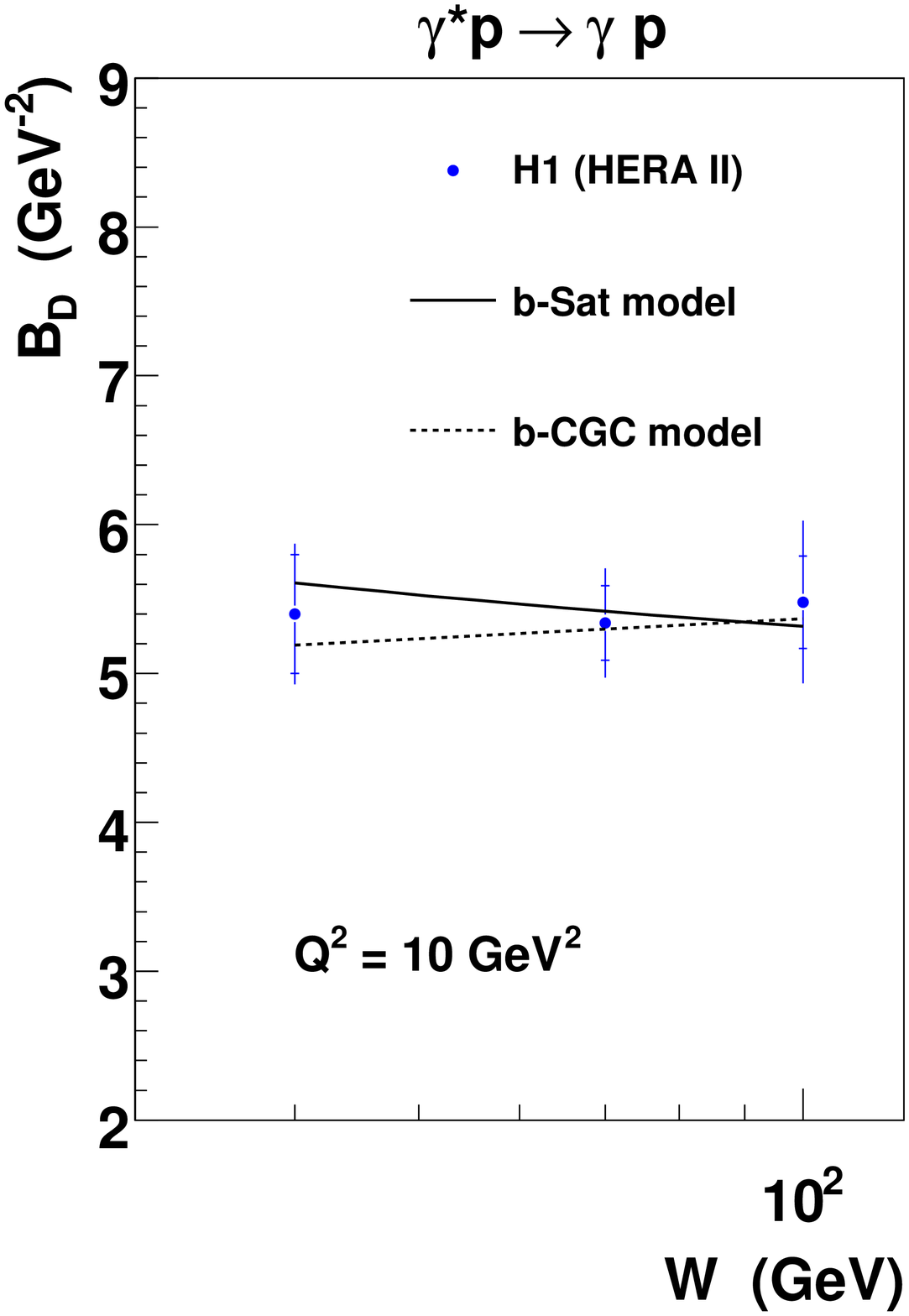}
  \caption{(a) The differential DVCS cross section $\dif{\sigma}/\dif{t}$ vs.~$|t|$ for different $Q^2$, and (b) the corresponding $t$-slope parameter $B_D$ vs.~$Q^2$, where $B_D$ is defined by fitting $\dif\sigma/\dif t\propto \exp(-B_D|t|)$.  Similarly, we show (c) $\dif{\sigma}/\dif{t}$ vs.~$|t|$ for different $W$, and (d) the corresponding $t$-slope parameter $B_D$ vs.~$W$.  The DVCS data \cite{Aktas:2005ty,Chekanov:2003ya,Aaron:2007cz,ZEUSDVCS} are compared to predictions from the b-Sat and b-CGC models.}
  \label{fig:dvcsdiff}
\end{figure}

In Fig.~\ref{fig:crossq} we show the description of the $Q^2+M_V^2$ dependence of the vector meson data.  In Fig.~\ref{fig:crossw} we show the $W$ dependence, and in Fig.~\ref{fig:delta} we show the $Q^2$ dependence of $\delta$, where $\sigma\propto W^\delta$.  In Fig.~\ref{fig:bd} we show the $Q^2$ dependence of $B_D$, where $\dif\sigma/\dif t\propto \exp(-B_D|t|)$.  In Fig.~\ref{fig:r} we show $R\equiv\sigma_L/\sigma_T$ vs.~$Q^2$, $W$ and $t$, for $\rho$ meson electroproduction.  In Fig.~\ref{fig:apom} we show the effective Pomeron trajectory $\alpha_\Pom(t)$ vs.~$|t|$, where $\alpha_\Pom(t)$ is determined by fitting $\dif\sigma/\dif t\propto W^{4[\alpha_\Pom(t)-1]}$.  In Fig.~\ref{fig:bdw} we show the $t$-slope parameter $B_D$ vs.~$W$.  Finally, in Figs.~\ref{fig:dvcstotal} and \ref{fig:dvcsdiff} we show the $Q^2$, $W$ and $t$ dependences for DVCS.

In general, both the b-Sat and b-CGC dipole models, using the ``boosted Gaussian'' vector meson wave function, describe almost every feature of the available data.  The b-CGC model gives a better description of $\alpha_\Pom^\prime$, where $\alpha_\Pom(t) = \alpha_\Pom(0)+\alpha_\Pom^\prime\,t$, as noted in Ref.~\cite{Kowalski:2006hc}, as seen from the slope of the predictions in Figs.~\ref{fig:apom} and \ref{fig:bdw}.  One major problem, again as noted in Ref.~\cite{Kowalski:2006hc}, is with the ratio $R\equiv\sigma_L/\sigma_T$ for $\rho$ meson electroproduction at large $Q^2$; see Fig.~\ref{fig:r}.  The flat behaviour of $R$ as a function of $|t|$ seen in the data implies that the $t$-slope parameter $B_D$ is the same for both transverse and longitudinal photon polarisations.  This surprising behaviour contradicts both the model predictions and the general intuitive picture, where the transversely polarised cross section is dominated by larger dipole sizes than the longitudinally polarised cross section, leading to $B_{D,T}>B_{D,L}$.  A detailed examination of the transversely polarised $\rho$ meson wave function is needed to resolve this issue.

\begin{figure}
  \begin{minipage}{0.5\textwidth}
    (a)\\
    \includegraphics[width=\textwidth,clip]{ampelT_r.eps}
  \end{minipage}%
  \begin{minipage}{0.5\textwidth}
    \hspace*{0.1\textwidth}(b)\\
    \includegraphics[width=\textwidth]{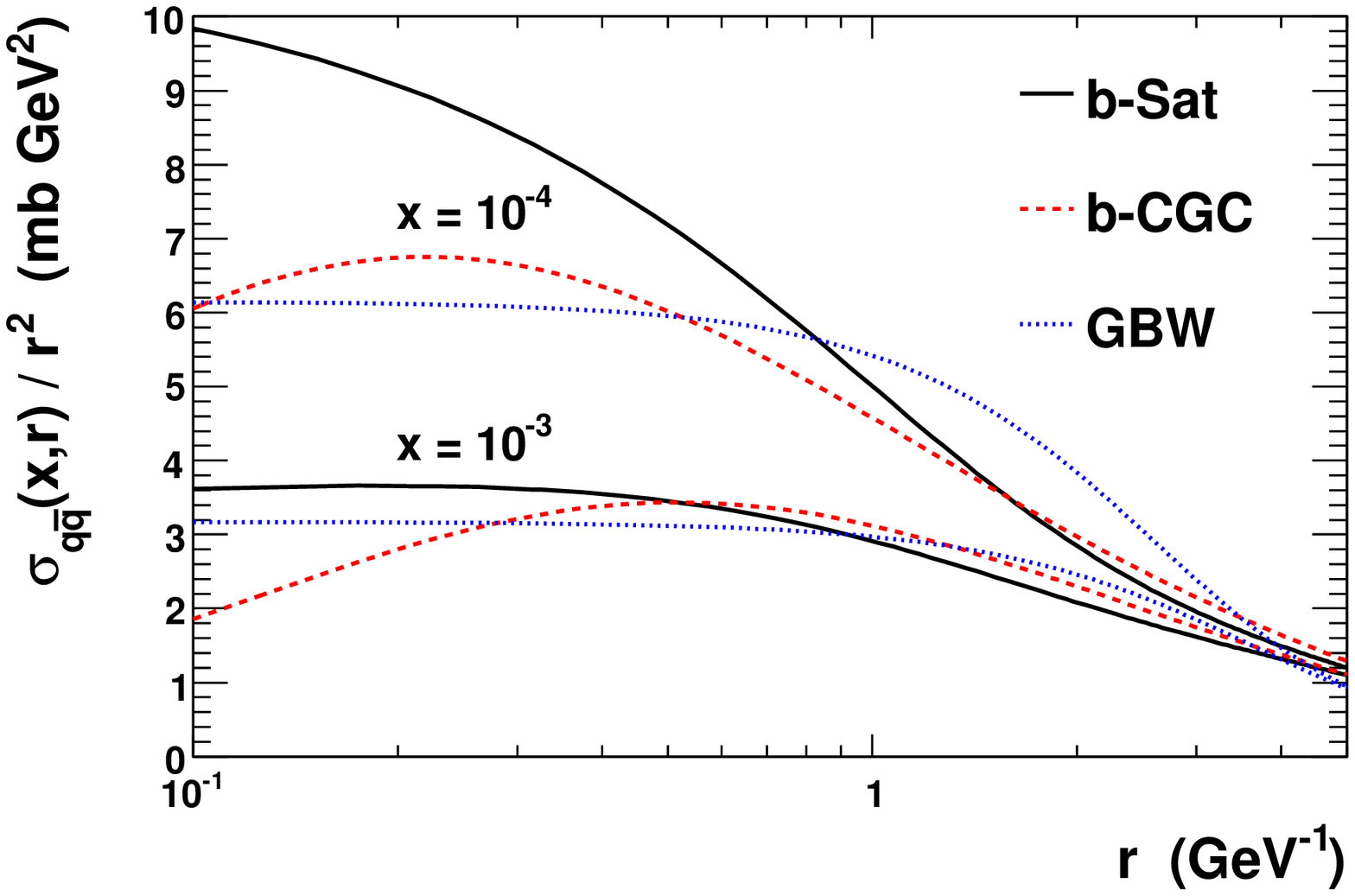}\\
    \hspace*{0.1\textwidth}(c)\\
    \includegraphics[width=\textwidth]{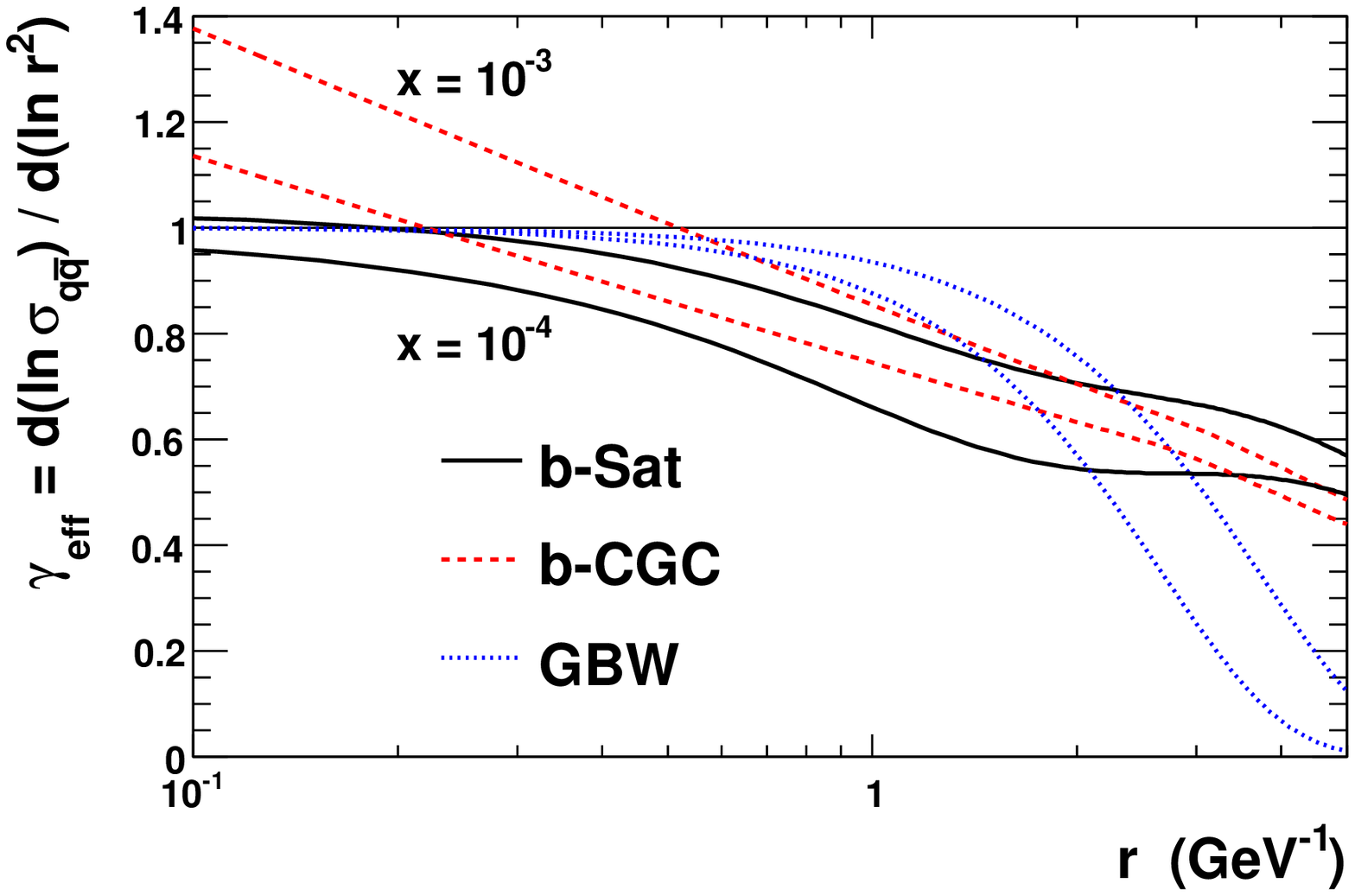}
  \end{minipage}
  \caption{(a) The $r$ dependence of the amplitude for exclusive $J/\psi$ photoproduction at $t=0$ (temporarily omitting the skewedness and real part corrections), (b) the $b$-integrated dipole cross sections divided by $r^2$, and (c) the effective anomalous dimension, $\gamma_{\rm eff}=\partial \ln\sigma_{q\bar{q}}/\partial \ln r^2$.}
  \label{fig:ampelT}
\end{figure}
The $W$ dependence of $J/\psi$ photoproduction is much better described by the b-Sat model than by the b-CGC model.  (The description with the b-CGC parameters given in the fourth line of Table \ref{tab:bcgc}, where high-$Q^2$ $F_2$ data were also included, is not any better.)  Indeed, the $W$ dependence, or the power $\delta$, is a test of the evolution of the dipole cross section (or generalised gluon density) and provides a powerful discriminator between the different models.  In Fig.~\ref{fig:ampelT}(a) we show the $r$ dependence of the imaginary part of the scattering amplitude for $J/\psi$ photoproduction.  The distributions are peaked at $r\approx 1.1$--$1.3$ GeV$^{-1}$, cf.~the ``scanning radius'' \cite{Nemchik:1994fp} of $r\approx 6/M_{J/\psi} = 1.9$ GeV$^{-1}$.  Going from $W=50$ GeV ($x=4\times10^{-3}$) to $W=300$ GeV ($x=10^{-4}$), the b-CGC model amplitude goes from being larger than the b-Sat model amplitude around the peak to being smaller.  This behaviour can be traced to the $b$-integrated dipole cross section in the relevant $r$ region, shown in Fig.~\ref{fig:ampelT}(b) at two different values of $x$.  For comparison, we also show the GBW fit, including charm quarks, from Ref.~\cite{Kowalski:2006hc}.  To examine the differences between the three parameterisations of the dipole cross section in more detail, we can also compute the effective anomalous dimension, $\gamma_{\rm eff}=\partial \ln\sigma_{q\bar{q}}/\partial \ln r^2$, shown in Fig.~\ref{fig:ampelT}(c).  For small dipole sizes, the value of $\gamma_{\rm eff}$ for the b-Sat (GBW) dipole cross section approaches a limiting value of approximately (exactly) 1.  On the other hand, the value of $\gamma_{\rm eff}$ for the b-CGC model diverges for small $r$ and large $x$; see \eqref{eq:anomdim}.

Although there are relatively small differences between the b-Sat and b-CGC dipole cross sections shown in Fig.~\ref{fig:ampelT}(b) around $r\sim 1$ GeV$^{-1}$, the $J/\psi$ data can discriminate between them, as seen more clearly in the description of the $\delta$ parameter for $J/\psi$ photoproduction shown in Fig.~\ref{fig:delta}.  Note that the skewedness and real part corrections in \eqref{eq:xvecm1}, not included in Fig.~\ref{fig:ampelT}, enhance the relative difference between the b-Sat and b-CGC dipole cross sections by about a factor $1.2$ due to the larger values of $\lambda$ obtained in the b-Sat model.  A similar difference between the b-Sat and b-CGC model predictions is found in the values of $\delta$ for DVCS shown in Fig.~\ref{fig:dvcstotal}; however, the HERA data on DVCS are not yet precise enough to distinguish between the two models.

We conclude that the b-CGC model, and other similar models derived from \eqref{eq:cgc}, work well for the designed purpose of extracting the saturation scale and describing low-to-moderate $Q^2$ DIS, and also for describing exclusive diffractive $\rho$ and $\phi$ meson production and DVCS.  However, such models fail for observables sensitive to relatively small dipole sizes, such as $J/\psi$ photoproduction and high-$Q^2$ DIS, where the b-Sat model \cite{Kowalski:2006hc} with explicit DGLAP evolution performs better.  Note that the failure of the b-CGC model to describe $J/\psi$ photoproduction does not indicate that the model should also fail for inclusive DIS at $Q^2 \sim M_{J/\psi}^2$, since exclusive diffractive processes are known to be dominated by smaller dipole sizes than inclusive DIS at the same $Q^2$; see, for example, Ref.~\cite{Kowalski:2003hm}.

\section{Conclusions} \label{sec:conclusions}

We have extended the CGC dipole model of Iancu, Itakura and Munier \cite{Iancu:2003ge} to include the impact parameter dependence.  The value of the anomalous dimension at the saturation scale, $\gamma_s=0.46$, determined from fitting to the total $\gamma^*p$ cross section at HERA, is close to the value of 0.44 obtained from numerical solution of the BK equation \cite{Boer:2007wf}.  However, the value of $\lambda=0.119$ is lower than might be expected from a perturbative calculation \cite{Triantafyllopoulos:2002nz}.  The impact parameter dependent saturation scale, $Q_S^2$, is generally $\lesssim 0.5$ GeV$^2$ in the HERA kinematic regime for the most relevant $b\sim 2$--$3$ GeV$^{-1}$.  This is in agreement with previous findings from the b-Sat model \cite{Kowalski:2006hc,Kowalski:2003hm}, that is, the Glauber--Mueller dipole cross section with DGLAP evolution of the gluon density.  We have shown that the data do not show a strong preference for the solution of the CGC model presented by Soyez \cite{Soyez:2007kg}, which has a large saturation scale $Q_S^2\sim 1$ GeV$^2$ at $x=10^{-5}$.  It would be interesting to check whether the present b-CGC dipole cross section is compatible with data from RHIC.  The dipole model predictions have been successfully compared with the first direct measurement of the longitudinal structure function from HERA \cite{Aaron:2008tx}, and also with existing HERA data on the charm and beauty structure functions.

Although a leading-twist collinear factorisation theorem has been proven \cite{Collins:1996fb} by which the amplitude for exclusive meson production can be expressed as a convolution of generalised parton distributions (GPDs) with hard-scattering kernels and meson distribution amplitudes, there are a number of practical problems with this approach (more progress has been made for DVCS \cite{Kumericki:2007sa}).  Firstly, the GPDs are not well-known and one needs to rely on models where, for example, the GPDs are written in terms of the usual parton distributions of the proton.  (However, the same criticism could be applied to the modelling of the dipole cross section.)  Secondly, the NLO corrections have been calculated and found to be huge at small $x$ \cite{Diehl:2007hd}, implying that small-$x$ resummation will be needed to achieve a stable result.  Therefore, the colour dipole approach discussed in this paper, or the related $k_t$-factorisation approach (see, for example, Refs.~\cite{Martin:1999wb,Ivanov:2003iy}), provide a complementary way to describe exclusive diffractive processes.  These approaches include some effects from small-$x$ resummation, and also corrections due to the transverse momentum of the partons entering the hard-scattering subprocess, which are neglected in the collinear factorisation approach but are known to be substantial \cite{Diehl:2007hd}.  Of course, the weakness of the dipole picture with respect to the collinear factorisation approach is that it is not currently known how to systematically improve the dipole picture beyond LO.

The b-CGC model presented in this paper has been shown to provide a good description of exclusive diffractive $\rho$ and $\phi$ meson electroproduction and DVCS, but performs less well for observables sensitive to relatively small dipole sizes, such as exclusive diffractive $J/\psi$ photoproduction and the beauty structure function at high $Q^2$.  In these cases, the b-Sat model \cite{Kowalski:2006hc} with explicit DGLAP evolution performs better.  The $W$ dependence of exclusive diffractive processes provides an important discriminator between different dipole model cross sections.  It would therefore be interesting to measure exclusive diffractive processes with greater precision than at HERA, as may be achievable at the LHeC \cite{Dainton:2006wd} or with a future electron--ion collider \cite{Deshpande:2005wd}.

It should be borne in mind that the dipole picture is not exact and that there are a number of assumptions and approximations made in its formulation \cite{Thorne:2005kj,Ewerz:2007md}.  Nevertheless, despite the de-emphasis of the saturation aspect of the dipole models, the framework provides an intuitive and economical description of a wide variety of small-$x$ processes.

\begin{acknowledgments}
  We thank Alan Martin, Leszek Motyka, Misha Ryskin and Robert Thorne for valuable discussions.  G.W.~acknowledges the UK Science and Technology Facilities Council for the award of a Responsive Research Associate position.
\end{acknowledgments}

\end{document}